\documentclass[12pt]{article}
\usepackage{hyperref} 
\usepackage{amsfonts}
\usepackage{amsmath, amssymb}
\usepackage{graphicx}
\usepackage{psfrag}

\newcommand{\ra}{\rightarrow}

\newcommand{\be}{\begin{equation}}
\newcommand{\ee}{\end{equation}}
\newcommand{\ba}{\begin{eqnarray}}
\newcommand{\ea}{\end{eqnarray}}
\newcommand{\bi}{\begin{itemize}}
\newcommand{\ei}{\end{itemize}}

\newcommand{\Tr}{{\rm Tr}}

\newcommand{\Z}{{\mathbb Z}}
\newcommand{\R}{{\mathbb R}}
\newcommand{\C}{{\mathbb C}}

\newcommand{\CP}{{\bf P}}

\newcommand{\Ncal}{{\mathcal N}}

\newcommand{\Scal}{{\mathcal S}}

\newcommand{\Acal}{{\mathcal A}}

\newcommand{\Zcal}{{\mathcal Z}}
\newcommand{\Kahler}{K\"{a}hler }

\newcommand{\nn}{\nonumber}
\newcommand{\mo}{{-1}} 

\newcommand{\f}{\frac}
\newcommand{\half}{\frac{1}{2}}
\newcommand{\oo}{\frac{1}}

\def\Dslash{\,\,{\raise.15ex\hbox{/}\mkern-12mu D}}
\def\Dbarslash{\,\,{\raise.15ex\hbox{/}\mkern-12mu {\bar D}}}
\def\delslash{\,\,{\raise.15ex\hbox{/}\mkern-9mu \partial}}
\def\delbarslash{\,\,{\raise.15ex\hbox{/}\mkern-9mu {\bar\partial}}}
\def\pslash{\,\,{\raise.15ex\hbox{/}\mkern-9mu p}}
\def\calDslash{\,\,{\raise.15ex\hbox{/}\mkern-12mu {\cal D}}}

\renewcommand{\bar}{\overline}

\begin{document}
\baselineskip=15.5pt
\renewcommand{\theequation}{\arabic{section}.\arabic{equation}}
\pagestyle{plain} \setcounter{page}{1}
\bibliographystyle{utcaps}
\begin{titlepage}

\rightline{\small{\tt EFI-07-32}}
\rightline{\small{\tt NSF-KITP-07-192}}
\begin{center}

\vskip 3 cm
\centerline{{\Large {\bf Bubbling Calabi-Yau geometry}}}
\vskip .5 cm 
\centerline{{\Large {\bf  from matrix models}}}

\vskip 2cm

{
Nick Halmagyi}

{\it Enrico Fermi Institute, University of Chicago

Chicago, IL 60637, USA
}
\vskip 7mm
and
\vskip 7mm
{
Takuya Okuda}

{\it Kavli Institute for Theoretical Physics, 
University of California

Santa Barbara, CA 93106, USA
}

\vskip 2cm

{\bf Abstract}

\end{center}
We study bubbling geometry in topological string theory. 
Specifically, we analyse Chern-Simons theory on both the 3-sphere and lens spaces in the presence of a
Wilson loop 
of an arbitrary representation. For each  three manifold,
 we formulate a multi-matrix model whose partition function 
is the Wilson loop vev
and compute the spectral curve. 
This spectral curve is closely related to 
the Calabi-Yau threefold which is 
the gravitational dual of the Wilson loop.
Namely, it is the reduction
to two dimensions of the mirror to the Calabi-Yau.
For lens spaces the dual geometries are new.
We comment on a similar matrix model
relevant for Wilson loops in AdS/CFT. 

\end{titlepage}

\newpage

\tableofcontents
\section{Introduction and summary}
A useful aspect of duality 
between a gauge theory and a gravitational system
is the emergence of spacetime through
dynamics of gauge theory.
Deeper understanding of emergent geometry
should help us find new formulations of
string theory and quantum gravity
that may be used to address fundamental questions
in physics.

In gauge/gravity duality, the vacuum state
corresponds to a certain
background spacetime, 
and inserted operators to excitations.
The fields of gauge theory
backreact significantly to the insertion of some operators. 
The corresponding gravitational dual is a new geometry
that shares the asymptotics 
with the original background.
A bubble of new cycles supported by flux appears,
and the new spacetime is thus called the {\it bubbling geometry}.
The bubbling phenomenon was originally found for local operators
 \cite{Lin:2004nb},
and was generalized to Wilson loops 
 \cite{Yamaguchi:2006te, Lunin:2006xr,  D'Hoker:2007fq}
in AdS/CFT.  
It is useful
to introduce a matrix model which captures
the dynamics of all the relevant fields
that respond to the operator insertion \cite{Berenstein:2004kk,Drukker:2000rr}.
One is able to visualize the backreaction    
in terms of eigenvalue distributions, which in turn
encode the bubbling geometry on the gravity side. 

\begin{figure}[htbb]
\begin{center}
\psfrag{n1}{$n_1$}
\psfrag{k1}{$k_1$}
\psfrag{n2}{$n_2$}
\psfrag{kg-1}{$k_{m-1}$}
\psfrag{ng}{$n_m$}
\psfrag{kg}{$k_m$}
\psfrag{ng+1}{$n_{m+1}$}
\includegraphics[width=80mm]{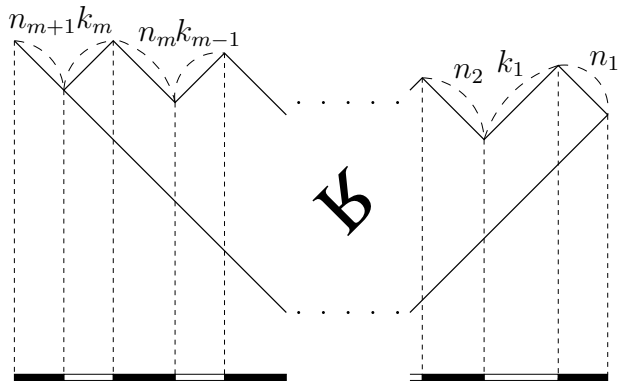} 
\caption{The Young tableau $R$, shown rotated and inverted, 
is specified by the lengths $n_I$ and $k_I$ of the
edges.
Equivalently, $n_I$ and $k_I$ denote the lengths 
of the black and white regions that
are obtained by vertically projecting down
the edges in $R$ onto the horizontal line.
$n_{m+1}$ is defined by $\sum_{I=1}^{m+1} n_I=N$.
}
\label{parametrization}
\end{center}
\end{figure}

The current work studies the topological string version of
 bubbling phenomena \cite{Gomis:2006mv},
which  naturally extend the Gopakumar-Vafa gauge/gravity duality  \cite{Gopakumar:1998ki}. 
More specifically we consider  $U(N)$ Chern-Simons
theory on $S^3$ or lens space $L(p,1)=S^3/\Z_p$ with Wilson loop insertions.
The Wilson loop operator is defined as
\ba
W_R\equiv \Tr_R e^{\oint A}
\ea
where $A$ is the gauge field and is integrated
along the unknot.
For $S^3/\Z_p$ we take the unknot that
generates the fundamental group.
The trace is
evaluated in an arbitrary representation $R$ of $U(N)$.
Throughout the paper the symbol $R$ also denotes the
corresponding Young tableau, and we parametrize it
as in Figure \ref{parametrization}.
Each edge length be it $n_I$ or $k_I$, will correspond to the
size of a new cycle in the bubbling geometry.

\begin{figure}[htbp]
\centering
\begin{tabular}{ccc}
\includegraphics[height=3cm]{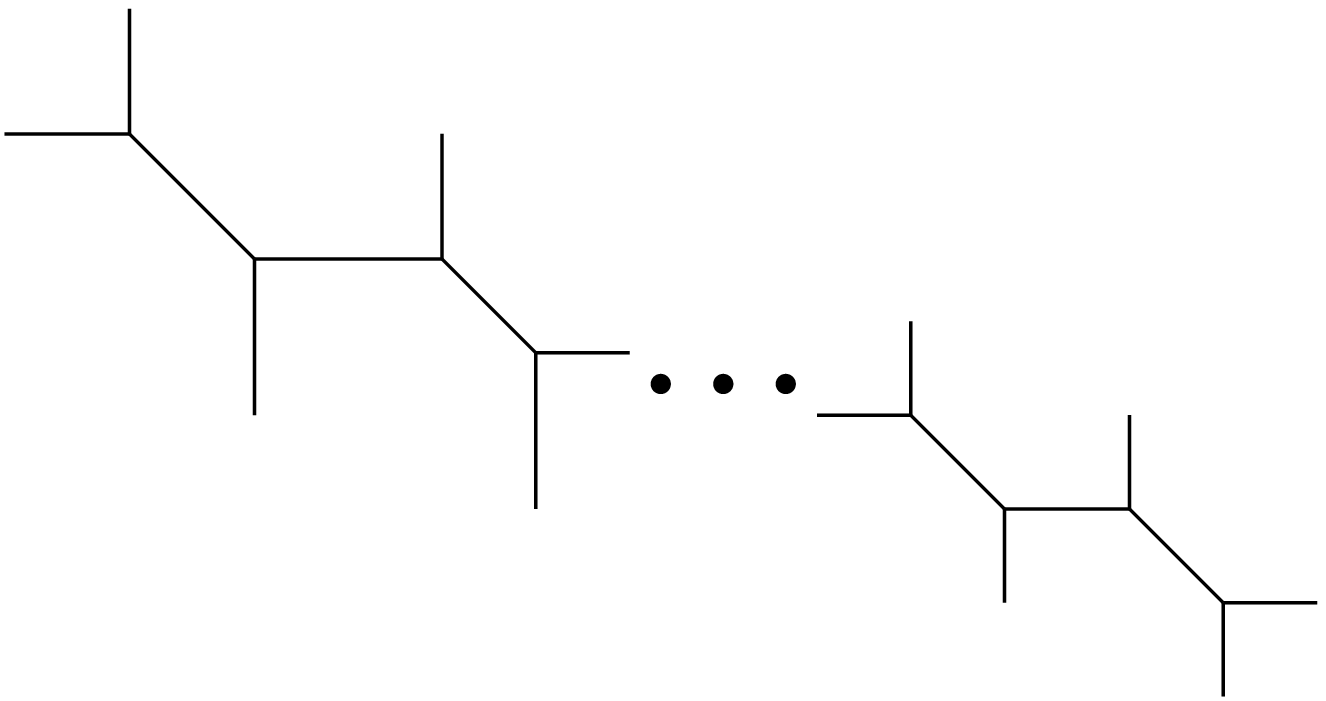}  
&
\includegraphics[height=3cm]{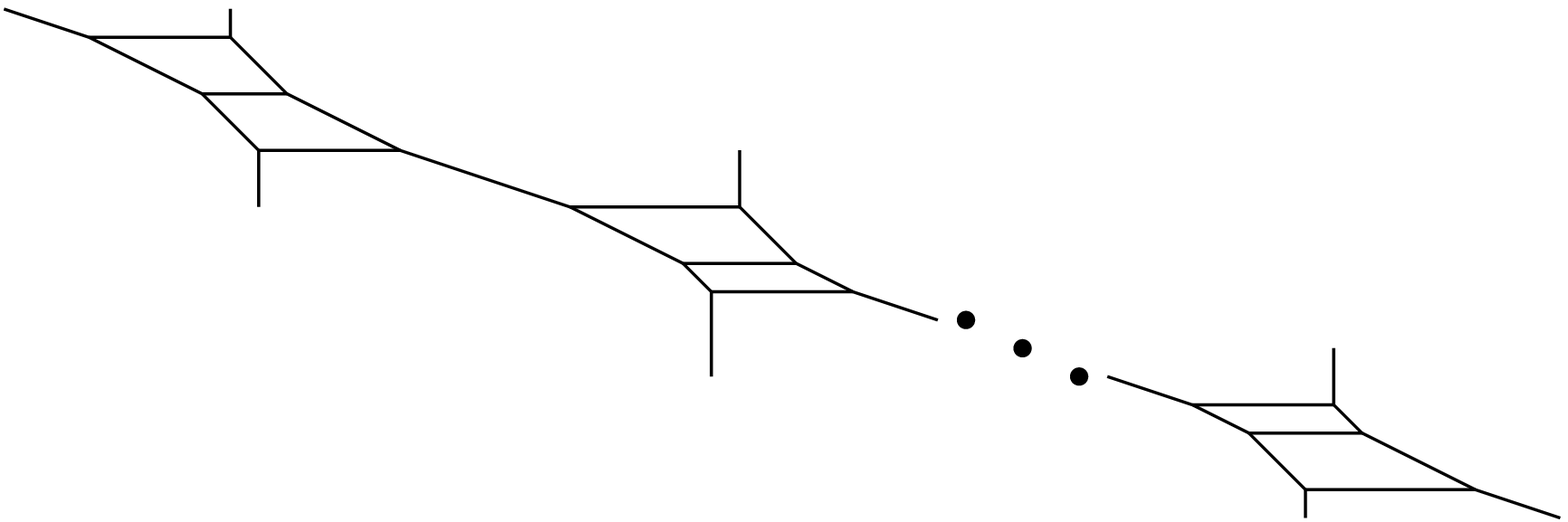}     
\\
(a)&(b)
\end{tabular}
\caption{(a) The toric web diagram for the bubbling Calabi-Yau dual to
the Wilson loop $W_R$ in $S^3$. 
It has $2m+1$ copies of $\CP^1$.
(b)
The web diagram for the bubbling Calabi-Yau dual to $W_R$ 
in lens space $S^3/\Z_p$ with $p=3$.
The diagram is a chain of $m+1$ basic units.
}
\label{bub-geom}
\end{figure}

Building on the earlier work \cite{Marino:2002fk,Aganagic:2002wv}, 
we formulate a matrix model whose partition
function is the vev of the Wilson loop in $S^3$ or $S^3/\Z_p$.
We then study the eigenvalue dynamics
in the large $N$ limit and derive the spectral curve.
For $S^3$ the spectral curve
is precisely the mirror of the
bubbling toric Calabi-Yau geometry identified
as the gravitational dual of the Wilson loop
in \cite{Gomis:2006mv}. The topology of this threefold depends on the data encoded in the Young tableau $R$: its toric web diagram is shown in Figure \ref{bub-geom}(a)

For the lens spaces $S^3/\Z_p$, the backreaction of the fields
to the Wilson loop
leads to additional classical vacua, and the 
path-integral splits into sectors corresponding to the different vacua.
Because the matrix model we formulate computes the Wilson loop vev
in each sector,
we  propose that
for given $N, p$, and $R$, a single Wilson loop insertion
is dual to a sum over bubbling geometries. Each term in the sum is the
toric Calabi-Yau that is mirror to the spectral curve which we derive. 
The summed geometries have the same toric data shown
in Figure \ref{bub-geom}(b)\footnote{
To take the limit $p\ra 1$ in Figure \ref{bub-geom},
one need to apply an $SL(2,\Z)$ transformation.
}
except different values of \Kahler moduli.
 As in the $S^3$ case, the topology of the geometry depends on the Young tableau data.

The paper is organized as follows.
Section \ref{S3section} focuses on the $S^3$ case.
In subsection \ref{S3matrix} we present
the matrix model for a Wilson loop in $S^3$.
In subsection \ref{phys-der-S3} we derive
the matrix model from physical arguments.
Specifically we 
present it as an open string field theory of a D-brane configuration
that realizes the Wilson loop.
Then we algebraically derive the matrix model
in subsection \ref{alg-der-S3}. 
In subsection \ref{S3spect}, we solve the matrix model
in the large $N$ limit and derive the spectral curve,
which is the mirror of the bubbling Calabi-Yau found in \cite{Gomis:2006mv}.

Section \ref{lenssection} deals with lens space $S^3/\Z_p$, 
and is structured in parallel with section \ref{S3section}.
For each vacuum of the gauge theory with Wilson loop insertion,
we derive the spectral curve.
We propose that the mirror toric Calabi-Yau is the
bubbling geometry dual to the Wilson loop.

Appendix \ref{Y-summary} summarizes the notation
regarding the Young tableau data.
In appendix \ref{S3app} we study alternative
matrix models that compute the Wilson loop vev.
The models are the direct analog of the matrix models
for $\Ncal=4$ Yang-Mills considered in \cite{Okuda:2007kh}.
Appendix \ref{YMsec} is targeted at readers interested in
AdS/CFT.
We use the algebraic techniques in subsection \ref{alg-der-S3}
to formulate a matrix model,
whose partition function is
the vev of the supersymmetric circular Wilson loop
in $\Ncal=4$ Yang-Mills.
In this formulation it is very easy to derive the
eigenvalue distributions for the Wilson loop
found in \cite{Hartnoll:2006is, Yamaguchi:2007ps,
 Yamaguchi:2006tq,Okuda:2007kh}.


\section{Bubbling Calabi-Yau for \texorpdfstring{$S^3$}{S3} from a matrix model}
\label{S3section}


\subsection{Matrix model for a Wilson loop in \texorpdfstring{$S^3$}{S3}}
\label{S3matrix}

The realization that the open topological A-model can be reduced to a matrix model first appeared in Marino's work \cite{Marino:2002fk}, and a B-model version of this idea was subsequently derived by Dijkgraaf and Vafa \cite{Dijkgraaf:2002fc}. Both derivations are of course mirror to each other as was demonstrated for certain examples in the nice work \cite{Aganagic:2002wv}. We are interested here in the A-model, which is of course equivalent to Chern-Simons theory \cite{Witten:1992fb}, possibly with instanton corrections \cite{Witten:1992fb, Diaconescu:2002sf, Aganagic:2002qg}.

Marino's observation for Chern-Simons theory on $S^3$ with 
the gauge group $G$ was that the partition function is
\ba
\Zcal &=& \int d_H u e^{-\frac{1}{2g_s} \Tr u^2} \nn \\
&=& \int \oo{N!}\prod_{i=1}^N d u_i \prod_{i<j} 
\left(2 \sinh \frac{u_i-u_j}{2} \right)^2
e^{-\frac{1}{2g_s} \sum_i u_i^2}, \label{S3integral}
\ea
where  the topological string coupling constant
$g_s$ is identified with the Chern-Simons coupling constant,
$U=e^u$, and $d_H u$ is the Haar measure on $G$ with
unusual integration range.
On the second line we specialized to the case $G=U(N)$,
and each $u_i$ is integrated from $-\infty$ to $+\infty$.
This observation by itself may not be overwhelming since it is a reformulation of Witten's classic result for the partition function \cite{Witten:1988hf}.  The main utility is the generalization to different manifolds, where they carry topological data \cite{Marino:2002fk, Aganagic:2002wv, Beasley:2005vf, Garoufalidis:2006ew} and to Wilson loops which we describe in the current work.

One very interesting feature of (\ref{S3integral}) however is that it secretly knows about the geometric transition of Gopakumar and Vafa \cite{Gopakumar:1998ki}. While Chern-Simons theory is equivalent to the open A-model on the {\it deformed} conifold, the spectral curve of (\ref{S3integral}) is directly related to the {\it resolved} conifold. 
If the Calabi-Yau threefold mirror to the resolved conifold is defined by
the equation $xy=f(e^u, e^v)$, the spectral curve
is then given by $f(e^u,e^v)=0$. The orientifold case was worked out in \cite{Halmagyi:2003fy}. Our main interest in this paper is to generalize this aspect to include the insertion of Wilson loop operators. Wilson loops in the topological gauge/gravity duality have been considered before by Ooguri and Vafa \cite{Ooguri:1999bv}. The current work and the previous work \cite{Gomis:2006mv,Gomis:2007kz, Okuda:2007ai} extends this in two ways. Firstly, the full backreaction of the Wilson loop is taken into account, as explained in \cite{Gomis:2007kz} this means the Wilson loop vev can be expressed in terms of purely {\it closed} string enumerative invariants. Secondly we provide a dictionary for a single Wilson loop in a particular representation $R$, whereas in \cite{Ooguri:1999bv} a sum of Wilson loop insertions  was considered where the summation is over representations.

Wilson loop operators
\ba
W_R=\Tr P e^{\oint A}
\ea
are specified by two pieces of data: the representation $R$ of the gauge group $G$ and a curve $\gamma$ in $M$ which the gauge field is integrated over. We will be considering all representations of $U(N)$ such that the $n_I$
and $k_I$ in Figure \ref{parametrization} are large and our $\gamma$ will be the unknot.  

The relation between Chern-Simons theory and the matrix model
was extended to include Wilson loops in \cite{Aganagic:2002wv}:
\ba \label{wilsonloop2}
\left\langle W_R\right\rangle = \int d_H u  
e^{-\f{1}{2g_s}\Tr( u^2)} \Tr_R e^u.
\ea 
We will show that the vev of the Wilson loop is 
in fact the partition function of the following matrix model:
\ba
\left\langle
W_R\right\rangle
&=&
\int\prod_{I=1}^{m+1} d_H u^{(I)} e^{-\f{1}{2g_s}\Tr u^{(I)}{}^2}
e^{L_I\Tr u^{(I)}}
\prod_{I<J} \det\left(e^{u^{(I)}/2}\otimes e^{-u^{(J)}/2}-
e^{-u^{(I)}/2}\otimes e^{u^{(J)}/2}\right)
\nn\\
&=& 
 \int\prod_I\Bigg(\oo{n_I!}\prod_i du^{(I)}_i
 \prod_{i<j} \left(2\sinh \f{u^{(I)}_i-u^{(I)}_j}2\right)^2
e^{-\f{1}{2g_s} \sum_i (u^{(I)}_i)^2}
e^{L_I\sum_i u^{(I)}_{i}}
\Bigg) \nn\\
&&
~~\times\prod_{I<J}\prod_{i,j}
\left(2\sinh\f{u^{(I)}_i-u^{(J)}_j}2\right),\label{S3model}
\ea
with 
\ba
L_I\equiv \sum_{J=I}^{m} k_J -\half\sum_{J=1}^{I-1}n_J
+\half \sum_{J=I+1}^{m+1}n_J~~
\hbox{ for }I=1,\ldots,m+1. \label{LI}
\ea
This is a Gaussian $(m+1)$-matrix model with certain interactions
which in the next section we explain from the target space viewpoint.


\subsection{Physical derivation of the matrix model} \label{phys-der-S3} 

\begin{figure}[htbp]
\centering
\begin{tabular}{ccc} 
\psfrag{R}{
}
\psfrag{N}{$N$}
\psfrag{N1}{$P$}
\includegraphics[scale=.45]{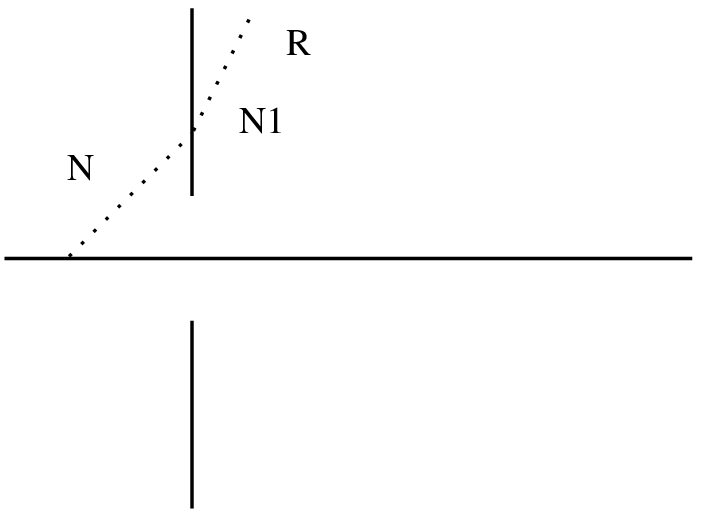}
&
\psfrag{nm+1}{$n_{m+1}$}
\psfrag{n2}{$n_2$}
\psfrag{nm}{$n_m$} 
\psfrag{n1}{$n_1$}
\includegraphics[scale=.45]{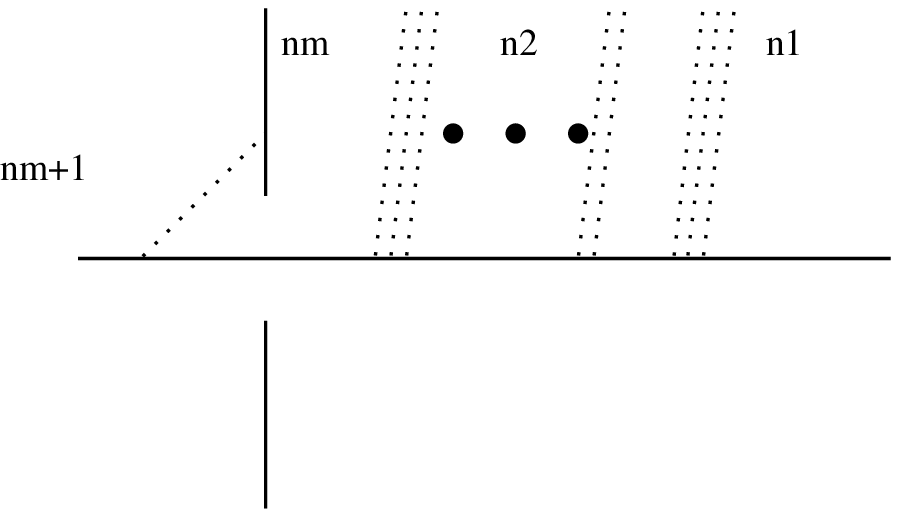}
&
\psfrag{nm+1}{$n_{m+1}$}
\psfrag{n2}{$n_2$}
\psfrag{nm}{$n_m$} 
\psfrag{n1}{$n_1$}
\includegraphics[scale=.45]{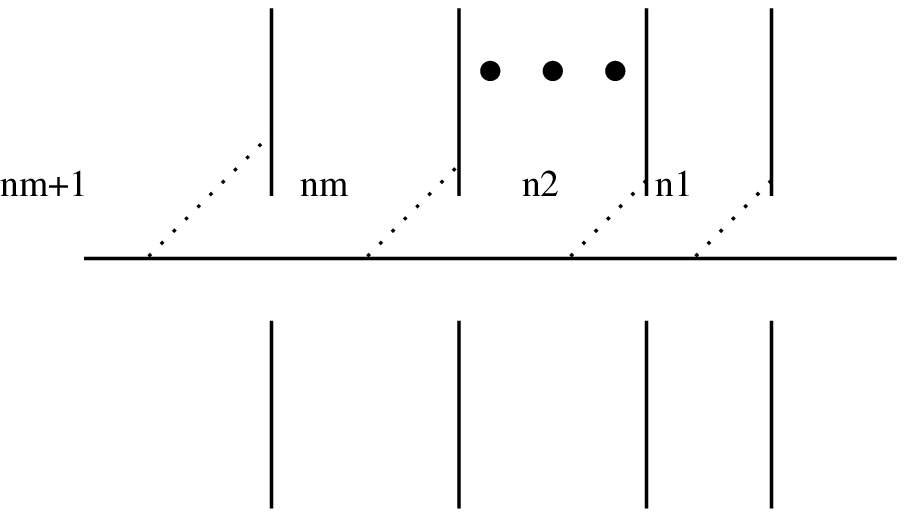}
\\
(a)&(b)&(c)
\end{tabular}
\caption{(a) The web diagram for the deformed conifold.
$\alpha$ and $\beta$ degenerate
along the horizontal and vertical lines respectively.
The dashed line represents $S^3$ that $N$ D-branes wrap.
The other dashed line ending on the vertical solid line
represents a non-compact cycle $L=\R^2\times S^1$
that $P$ non-compact D-branes wrap.
(b)
$P$ non-compact branes are distributed along the horizontal
line where $\alpha$ degenerates.
}
\label{physical1}
\end{figure}

In this subsection we derive the matrix model (\ref{S3model})
as the world-volume theory in a D-brane configuration that
is equivalent to the Wilson loop insertion.
Further geometric transition of the branes leads
to the purely closed string geometry in Figure \ref{bub-geom}(a),
and the three steps are summarized in Figure \ref{physical1}.
As we will describe, the essential details in each step can be found in the earlier work \cite{Gomis:2006mv,Gomis:2007kz, Okuda:2007ai}. 

We start with the deformed conifold geometry given by the equation
\ba
z_1 z_2=w=z_3 z_4+\mu,~~z_i, w\in \C, \label{def-con-eq}
\ea
where $\mu$ is the complex structure parameter
that we take to be real positive.
The geometry has
the structure of $T^2\times \R$ fibration over $\R^3$.
Let us denote the basis cycles of $T^2$ by $\alpha$ and $\beta$.
In the base $\R^3$, $\alpha$ degenerates along one line
and  $\beta$ degenerates on another.
The minimal $S^3$ is obtained by fibering this $T^2$ along a line interval
that connects the two loci.

We wrap $N$ branes on $M= S^3$ thus engineering the $U(N)$ 
Chern-Simons theory.
In addition we place a stack of $P$ branes\footnote{
Recall that $P$ is the number of rows in $R$.}
 wrapping a 
non-compact three cycle $L$ of topology ${\mathbb R}^2\times S^1$.
The cycle $L$ intersects $M$ along a circle that is identified with $\alpha$.
These branes were introduced in \cite{Ooguri:1999bv} where 
the partition function obtained after integrating out the 
bifundamental $M$-$L$ strings was shown to be 
a generating function for Wilson loop vevs.
The generating function is a summation over representations 
of $U(N)$ and the $S^1$ common to $M$ and $L$ is the 
defining curve of the Wilson loop. 
Since $L$ is non-compact 
one should enforce a boundary condition at infinity 
for the gauge field on the stack of branes which wrap $L$. 
In \cite{Ooguri:1999bv} this was implicitly done by fixing
the background holonomy of the gauge field along $\alpha$.

A different boundary condition
isolates a single Wilson loop in the representation
$R$ \cite{Gomis:2007kz}.
So this brane construction
is equivalent to the Wilson loop insertion.
See Figure \ref{physical1}(a).
This boundary condition is equivalent to the gauge field having a 
nontrivial holonomy matrix  along the $\beta$
 cycle which encodes the data of $R$.

First, the above brane configuration
is
equivalent to
another 
system
 that has a new set of non-compact D-branes,
distributed along the locus where $\alpha$ degenerates
\cite{Okuda:2007ai}. 
The new system has only $N-P(=n_{m+1})$ D-branes wrapping the $S^3$.
As we review in Appendix \ref{appendix-area},
a stack of $n_I$ non-compact branes 
sits at distance $a_I=g_s(L_I-L_{m+1})$ away from the $S^3$
for $I=1,\ldots, m$.
See Figure \ref{physical1}(b).

Second, by considering the new 
ambient geometry of Figure \ref{physical1}(c) with 
more complex structure moduli given by
\ba
z_1 z_2=w,~~~ z_3 z_4=(\mu-w)\prod_{I=1}^m(1-w/\mu_I),\label{new-geom-eq}
\ea
the non-compact branes can be compactified 
without changing the physics.
This is a legitimate maneuver since it reduces to the deformed conifold
 (\ref{def-con-eq})
by making the complex structure moduli $\mu_I$ infinite and
A-model depends only on \Kahler moduli. 
The result is the D-brane system from which we can derive the matrix model
(\ref{S3model}).

We now have a daisy chain of Chern-Simons theories 
all of them on an $S^3$ and there are then 
annulus instantons which connect them \cite{Aganagic:2002qg}. 
The representation $R$ of the Wilson loop determines all the
necessary data, in particular
the $I$-th Chern-Simons theory has
gauge group $U(n_I)$, $I=1,\ldots,m+1$.
We get annulus instantons 
by integrating out the massive bifundamental
open strings \cite{Ooguri:1999bv}.
Since the 
mass of
the string between
the $I$-th and the $J$-th spheres
is $a_I-a_J$,
the interactions generated from such annulus instantons
are summarized as 
\ba
\langle W_R\rangle\sim
\int \prod_{I=1}^{m+1}[DA_I]e^{iS_{CS}(A_I)}
\prod_{I<J}
\det\left(
e^{\half (a_I-a_J)}
U_I^{\half} \otimes U_J^{-\half}-
e^{\half (a_J-a_I)}
U_I^{-\half} \otimes U_J^{\half}\right).
\label{path-int2}
\ea
where $S_{CS}$ is the Chern-Simons action, and
$U_I\equiv P \exp\oint_\alpha A_I$ is the
holonomy along  the unknot in the $I$-th $S^3$.
Given this field theory description, we can now reduce
it to a matrix model \cite{Aganagic:2002wv}:
\ba
\hspace{-5mm}
\langle W_R\rangle
\hspace{-1mm}
\sim
\hspace{-1mm}
\int\prod_{I=1}^{m+1} d_H u^{(I)} e^{-\f{1}{2g_s}\Tr (u^{(I)})^2}
\prod_{I<J}
\det
\left(
2\sinh\f{(a_I+u^{(I)})\otimes 1- 1\otimes (a_J+u^{(J)})}2
\right).
\ea
By redefining $u^{(I)}\ra u^{(I)}-g_s L_I=u^{(I)}-a_I+(I$-independent), 
we finally obtain (\ref{S3model}).
It is a nontrivial consistency check that 
the physical derivation here gives the values of
holonomy $a_I$ that we need to
agree with the algebraic derivation in the next
subsection.

Third and finally,
we can go one step further in the target space analysis
though we have completed our task in this subsection,
When each $S^3$ in Figure \ref{physical1}(c) 
undergoes a conifold transition the resulting closed string geometry 
is the toric Calabi-Yau manifold whose web diagram is shown in 
Figure \ref{bub-geom}(a). 
This is the Calabi-Yau manifold which is 
referred to as the {\it bubbling} geometry \cite{Gomis:2006mv}.  
We will see in subsection \ref{S3spect} that the eigenvalue dynamics
in the matrix model
demonstrates the geometric transition.


\subsection{Algebraic derivation of the matrix model}\label{alg-der-S3}

We now provide an algebraic derivation of (\ref{S3model}). 
Our starting point is (\ref{wilsonloop2}).
Using a standard formula for the character of $U(N)$ this can be written as
\be
\left\langle W_R\right\rangle =
\int \oo{N!}\prod_i du_i \prod_{i<j} \left(2\sinh \f{u_i-u_j}2\right)^2
e^{-\f{1}{2g_s}\sum u_i^2}
\f{\det( e^{(N+R_j-j)u_i})}{\det (e^{(N-j)u_i})},
\ee
where $R_j$ is as usual the number of boxes in the $j$-th row of $R$.
Now we expand this ratio of determinants into something more compatible with the matrix model:
\ba
\f{\det( e^{(N+R_j-j)u_i})}{\det (e^{(N-j)u_i})}
&=&
\sum_{\sigma\in {\cal S}_N}{\rm sgn}(\sigma) 
\prod_i e^{(N+R_i-i)u_{\sigma(i)}}/\prod_{i<j}(e^{u_i}-e^{u_j})\nn \\
&=&
\sum_{\sigma\in {\cal S}_N}
\prod_i e^{(N+R_i-i)u_{\sigma(i)}}/\prod_{i<j}(e^{u_{\sigma(i)}}-e^{u_{\sigma(j)}}).
\ea
Since $u_i$ are dummy variables the summation over the permutation group
${\cal S}_N$ 
produces $N!$ identical terms, so we can write
\ba
\left\langle
W_R\right\rangle
&=&
\int \prod_i du_i \prod_{i<j} \left(2\sinh \f{u_i-u_j}2\right)^2
e^{-\f{1}{2g_s}\sum u_i^2}
\prod_i
e^{(N+R_i-i)u_{i}}/\prod_{i<j}(e^{u_{i}}-e^{u_{j}})\nn.
\ea
At this point the Wilson loop insertion has been recast into a linear term in the exponential and a certain denominator term. There will be partial cancellation of this denominator term against the measure and also against the linear term.
We relabel the variables as
\ba
(u_1,\ldots,u_N)=(u^{(1)}_1,\ldots,u^{(1)}_{n_1},
u^{(2)}_1,\ldots,u^{(2)}_{n_2},
\ldots,
u^{(m+1)}_1,\ldots,u^{(m+1)}_{n_{m+1}}) \nn
\ea
where we recall that the Young tableau $R$ has $m$ blocks of rows.
Then
\ba
\left\langle W_R\right\rangle
&=&
\int \prod_{I=1}^{m+1} \prod_{i=1}^{n_I} du^{(I)}_i
\prod_I \prod_{i<j} \left(2\sinh \f{u^{(I)}_i-u^{(I)}_j}2\right)^2
\prod_{I<J}\prod_{i,j}
\left(2\sinh \f{u^{(I)}_i-u^{(J)}_j}2\right)^2\nn \\
&&
\times\,\, e^{-\f{1}{2g_s}\sum_{I,i}  (u^{(I)}_i)^2}
\prod_I\prod_i
e^{ (N+K_I-(N_{m-I+2}+i))u^{(I)}_{i}} \nn \\
&& \times \left(
\prod_I \prod_{i<j}(e^{u^{(I)}_{i}}-e^{u^{(I)}_{j}})
\prod_{I<J}\prod_{i,j}(e^{u^{(I)}_{i}}-e^{u^{(J)}_{j}})\right)^{-1}.\nn
\ea
The integers $K_I$ and $N_I$ are defined in Appendix \ref{Y-summary}.
This can be further simplified,
using the trivial fact that integration variables are dummy variables,
to
\ba
&&\left\langle W_R\right\rangle
\nn\\
&=&
\hspace{-2mm}
 \int\prod_{I=1}^{m+1}\Bigg(\oo{n_I!}\prod_i du^{(I)}_i
 \prod_{i<j} \left(2\sinh \f{u^{(I)}_i-u^{(I)}_j}2\right)^2
e^{-\f{1}{2g_s} \sum_i (u^{(I)}_i)^2}
e^{(N+K_I-N_{m-I+2}-n_I)\sum_i u^{(I)}_{i}}
\nn\\
&&
~~\times \sum_{\sigma_I\in S_{n_I}}
e^{\sum_i (n_I-i) u^{(I)}_{\sigma_I(i)}}
/\prod_{i<j}(e^{u^{(I)}_{\sigma_I(i)}}-e^{u^{(I)}_{\sigma_I(j)}})
\Bigg)
\prod_{I<J}\prod_{i,j}
\frac{\left(2\sinh \f{u^{(I)}_i-u^{(J)}_j}2\right)^2}
{e^{u^{(I)}_{i}}-e^{u^{(J)}_{j}}}\nn
\\
&=&
 \int\prod_{I=1}^{m+1}\Bigg(\oo{n_I!}\prod_i du^{(I)}_i 
 \prod_{i<j} \left(2\sinh \f{u^{(I)}_i-u^{(I)}_j}2\right)^2
e^{-\f{1}{2g_s} \sum_i (u^{(I)}_i)^2} e^{K_I\sum_i u^{(I)}_{i}} \Bigg) \nn\\
&&
~~\times
\prod_{I<J}\prod_{i,j}\f{
\left(2\sinh \f{u^{(I)}_i-u^{(J)}_j}2\right)^2
}
{1-e^{u^{(J)}_{j}-u^{(I)}_{i}}}
\nn\\
&=&
 \int\prod_{I=1}^{m+1}\Bigg(\oo{n_I!}\prod_i du^{(I)}_i 
 \prod_{i<j} \left(2\sinh \f{u^{(I)}_i-u^{(I)}_j}2\right)^2
e^{-\f{1}{2g_s} \sum_i (u^{(I)}_i)^2} 
e^{L_I\sum_i u^{(I)}_{i}} \Bigg) \nn\\
&&~~\times
\prod_{I<J}\prod_{i,j}
\left(2\sinh \f{u^{(I)}_i-u^{(J)}_j}2\right),
 \label{WRfinal}
\ea
where $L_I$ are defined in (\ref{LI}) and we have use the relations
\ba
\prod_{I<J} \prod_{i,j} 
\left(2\sinh \f{u^{(I)}_i-u^{(J)}_j}2
\right)^2=e^{\sum_I (n_I-N)\sum_i u^{(I)}_i} \prod_{I<J} \prod_{i,j} (e^{u^{(I)}_i}-e^{u^{(J)}_j})^2,\nn\\
 \prod_{I<J} \prod_{i,j} (e^{u^{(I)}_i}-e^{u^{(J)}_j})
=e^{\sum_I(N-N_{m-I+1})\sum_i u^{(I)}_i} 
\prod_{I<J}\prod_{i,j}(1-e^{u^{(J)}_j-u^{(I)}_i}).
\ea
At this point we essentially have an $m$-matrix model with interactions between the matrices given by the last line in (\ref{WRfinal}).


\subsection{Spectral curve as the bubbling geometry 
dual to a Wilson loop in \texorpdfstring{$S^3$}{S3}}\label{S3spect}

Matrix models have an associated geometry
 called the spectral curve. One can think of $\langle W_R \rangle$ as a single Gaussian matrix model with somewhat complicated insertion, or alternatively as we have demonstrated, as an $m$-matrix model with certain simpler interactions. Taking the latter point of view, we now derive the spectral curve and explain its string theory interpretation.

The equations of motion for $u_{i}^{(I)}$ are
\be \label{S3eom}
0=-u_i^{(I)} +g_s L_I
+g_s \sum_{j\neq i}\coth \frac{u_i^{(I)}-u_j^{(I)}}{2} 
+\half g_s \sum_{J\neq I,i,j} \coth \frac{u_i^{(I)}-u_j^{(J)}}{2}. 
\ee
To solve them we define the resolvents\footnote{
The resolvents $\omega^{(I)}=
g_s \sum_{i=1}^{n_I} \coth \frac{z-u^{(I)}_i}{2}$
in another natural definition
are simply related to the $v^{(I)}$ as $\omega^{(I)}=g_s n_I-2v^{(I)}$.}
\be \label{resolvents}
v^{(I)}(z)=g_s \sum_{i=1}^{n_I}\f{e^{u^{(I)}_i}}{e^{u^{(I)}_i}-e^z},~~
v(z)= \sum_{I=1}^{m+1}v^{(I)}(z).
\ee
We now assume that the eigenvalues distribute themselves into $m$ distinct cuts along the real axis, then write (\ref{S3eom}) 
an equation 
on the $I$-th cut:
\ba
z+v^{(I)}_+(z) +v^{(I)}_-(z) +\sum_{J\neq I} v^{(J)}(z)
=g_s\left(\sum_{J=I}^m k_J+\sum_{J=I}^{m+1}n_J\right),
\label{eomprinc}
\ea
where $v^{(I)}_+(z)$ and $v^{(I)}_-(z)$ are the values
of $v^{(I)}(z)$ just above and below the cut, respectively.
It will be convenient to rewrite this as
\ba
z+v_{\pm}(z)=
-v^{(I)}_\mp(z) +g_s\left(\sum_{J=I}^m k_J+\sum_{J=I}^{m+1}n_J\right).
\label{S3eomb}
\ea

To derive the spectral curve, we generalize the 
complex analysis technique used in \cite{Halmagyi:2003ze} to solve the Chern-Simons matrix model
for $S^3/{\mathbb Z}_p$. The crucial step in solving this model is to find a set of functions of the resolvents $v^{(I)}$ which are regular on the whole
$Z$-plane where $Z=e^z$ is of course ${\mathbb C}^*$ valued. Then the asymptotics of $v^{(I)}$ will allow us to fix these functions exactly and finally extract the equation for the spectral curve. 
The technical reason that we will be able to solve this model exactly is that the interaction terms in the equation of motion can be written polynomially in terms of the resolvents. 
This is not the case for the related ${\cal N}=4$ Yang Mills matrix models 
described in appendix \ref{YMsec} and  also in \cite{Okuda:2007kh}. 

We first define some new quantities
\ba
X_{0}(Z)&=& Z e^v ,
\nn\\
X_I(Z)&=&A_Ie^{-v^{(I)}},\ \ \ I=1,\ldots,m+1,
\ea 
where $Z=e^{z}$ and $A_I=\exp g_s\left(\sum_{J=I}^m k_J+\sum_{J=I}^{m+1}n_J\right)$.
Equation (\ref{S3eomb}) implies that
$X_0$ and $X_I$ are exchanged as one goes through the $I$-th cut,
leaving any symmetric polynomial of $(X_0,X_1,\ldots,X_{m+1})$
invariant under the process.
The symmetric polynomial  is regular on all of the cuts, and
the only singularities are at $Z=\infty$.
Let us now recall the definition of the $j$-th 
{\it elementary symmetric polynomials} $E_j$:
\be
E_j(x_1,\ldots,x_n)=\sum_{i_1<\cdots<i_j} x_{i_1}\ldots x_{i_j}.
\ee
Together with the definition  (\ref{resolvents}) of the resolvents, 
the asymptotics as 
$z\rightarrow \pm \infty$ determine 
the $E_j(X_0,\ldots,X_{m+1})$ exactly in terms of Young tableau data:
\ba
E_0(X_0,\ldots,X_{m+1})&=&1,\nn\\
E_j(X_0,\ldots,X_{m+1})&=&a_{j,0}+a_{j,1}Z~~\hbox{ for }j=1,\ldots,m+1,
\label{Ejs}\\
E_{m+2}(X_0,\ldots,X_{m+1})&=&A_1\ldots A_{m+1}Z.\nn
\ea
The coefficients are given by
\ba
a_{j,0}&=&\sum_{1\leq J_1<\cdots<J_{j}\leq m+1} 
B_{J_1}
\ldots B_{J_j}
~~
\hbox{ for }j=1,\ldots,m+1,\nn\\
a_{j,1}&=&
\sum_{1\leq J_1<\cdots<J_{j-1}\leq m+1} A_{J_1}\ldots A_{J_{j-1}}~~\hbox{ for }
j=2,\ldots,m+1,
~~a_{1,1}\equiv 1, 
\ea
where we introduced 
$B_I=\exp g_s\left(\sum_{J=I}^m k_J+\sum_{J=I+1}^{m+1}n_J\right)$.

In fact the $E_j$ appear as the coefficients of $Y^j$ 
in the expansion of the function
\ba \label{detexpansion}
f(Y,Z)&\equiv&\prod_{J=0}^{m+1}(Y-X_J(Z))\nn\\
&=&\sum_{j=0}^{m+2}(-)^{j}Y^{m+2-j} E_{j}(X_0,\ldots,X_{m+1})
\nn\\
&=&
Y^{m+2}+
\sum_{j=1}^{m+1}(-1)^{j}Y^{m+2-j} (a_{j,0}+a_{j,1}Z)
+(-1)^{m_2} A_1\ldots A_{m+1}Z,
\ea
and this vanishes upon substituting $X_I$ for $Y$.
So we arrive at an equation for the spectral curve of the matrix model (\ref{S3model}):
\be
f(Y,Z)=0, \label{specurve}
\ee 
where  $(Y,Z)$ are ${\mathbb C}^*$ valued variables.

\begin{figure}[bth!]
\begin{center}
\begin{tabular}{ccc}
\includegraphics[scale=.25]{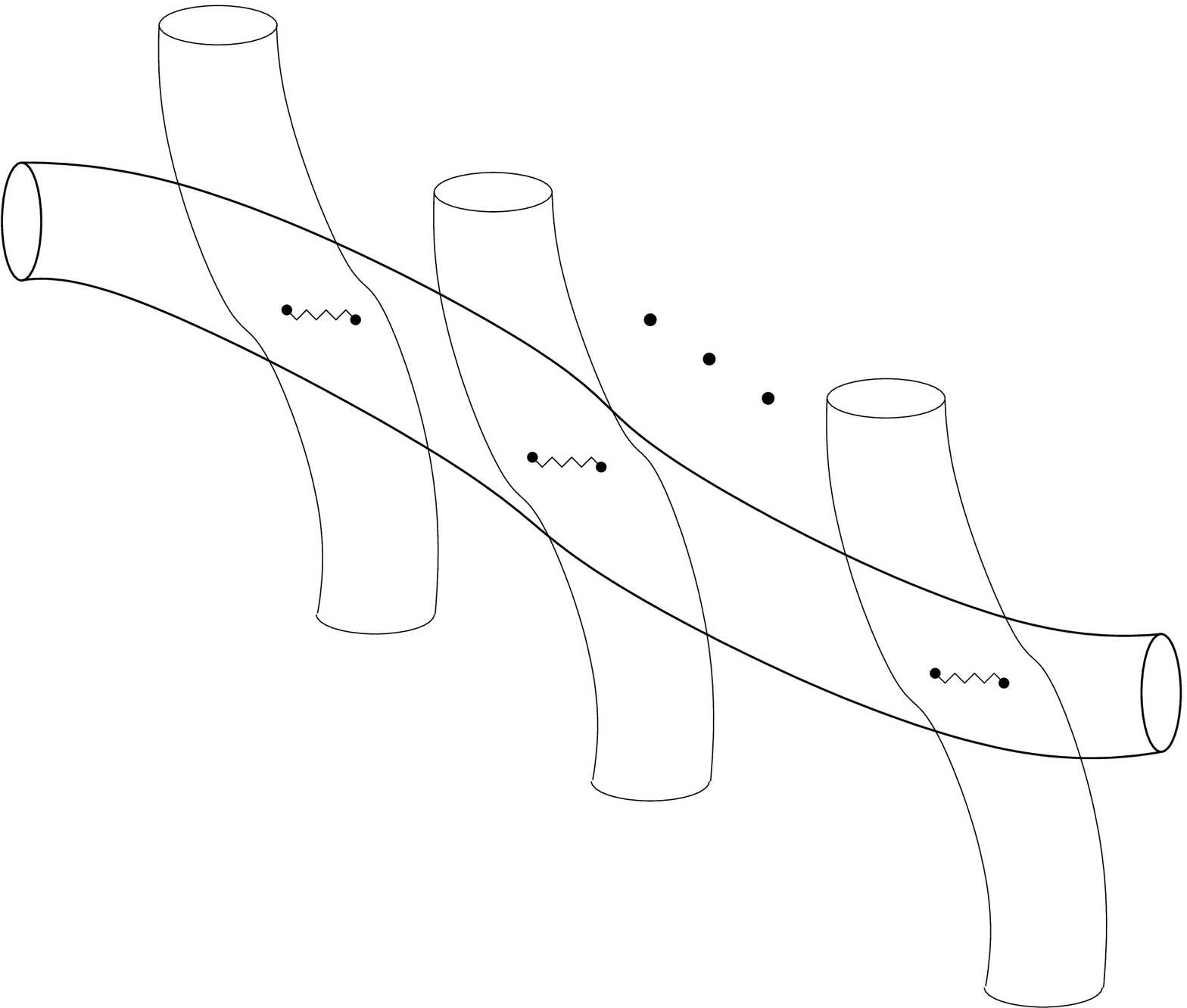} 
&
\hspace{1cm}
&
\psfrag{a}{$a$}
\psfrag{b}{$b$}
\includegraphics[scale=.8]{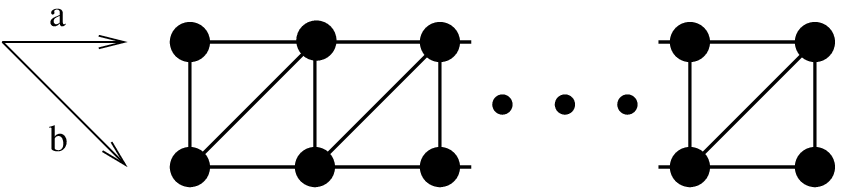}  
\\
(a)&&(b)
\end{tabular}
\caption{
(a)
The spectral curve is constructed by gluing one sheet
to $m+1$ other sheets through $m+1$ cuts.
Each sheet is a cylinder parametrized by $z$ with
identification $z\sim z+2\pi i$.
Compare with Figure \ref{bub-geom}(a).
(b) The vertices 
plot the monomials $Y^a Z^b$ appearing in the equation (\ref{specurve})
for the spectral curve.  By  connecting the vertices
by suitable edges, one obtains a graph that is dual to the toric
web for the bubbling geometry shown in Figure \ref{bub-geom}(a).
}
\label{snaky-S3}
\end{center}
\end{figure}

Since $f(Y,Z)$ is of degree $m+2$ in $Y$, the spectral curve is obtained by gluing $m+2$ cylindrical sheets.
In particular (\ref{specurve}) is satisfied by the total
resolvent $v(z)$ through substitution
$Y=X_{0}\equiv e^{z+v}$, 
and the sheet on which $v(z)$ is naturally defined has $m+1$ cuts.
By going through the $I$-th cut ($I=1,\ldots,m+1$), one
moves to the $I$-th sheet as $v(z)$ changes to
$-z-v^{(I)}+{\rm const.}$  See Figure \ref{snaky-S3}(a).

This Riemann surface is related to a Calabi-Yau threefold 
in a way which is by now well known, namely the threefold is given by
\be
wx=f(Y,Z)
\ee 
where $w,x$ are ${\mathbb C}$ valued. It is a feature of the mirror symmetry work of Hori-Vafa \cite{Hori:2000kt} that we can write down the toric fan directly from the Riemann surface data above. The recipe is to insert a vertex $(a,b)$ on the integral 2-dimensional
lattice for each monomial 
$Y^a Z^b$ appearing in (\ref{specurve}).
By connecting the vertices with suitable edges\footnote{
In the limit of large  $g_s n_I$ and $g_s k_I$
(the large volume limit in the $S^3$ case, but not in the $S^3/\Z_p$ case),
the difference between the GLSM algebraic coordinates \cite{Morrison:1994fr}
and our moduli is suppressed.
The mirror curve of the GLSM in this limit agrees with 
our spectral curve including the coefficients,
with the choice of internal edges in Figure \ref{snaky-S3}(b)
}  
one obtains a graph, and
the three-dimensional cone over this graph is
the toric fan of the bubbling Calabi-Yau.
The two dimensional graph is the dual graph of the
toric web diagram, so from Figure \ref{snaky-S3}(b)
we see agreement with the previous work \cite{Gomis:2006mv}.

For concreteness we now work out the simplest case when $R$ is a rectangle.
In this case the nontrivial data is 
\ba
A_1&=&e^{t_1+t_2+t_3},\ \ A_2= e^{t_3}, \nn\\
B_1&=&e^{t_2+t_3},\ \ B_2=1
\ea
and 
\ba
a_{1,0} &=& 1+ e^{t_2+t_3},\ \ a_{1,1}=1,  \\
a_{2,0}&=&e^{t_2+t_3},\ \ a_{2,1}= e^{t_1+t_2+t_3} + e^{t_3}, \nn \\
a_{3,0} &=& 0,\ \ a_{3,1}= e^{t_1+t_2+2t_3},\nn
\ea
with $t_1=g_s n_1,~t_2=g_s k_1,~t_3=g_s n_2$,
and so the spectral curve is explicitly given by
\ba
Y^3 -(1+ e^{t_2+t_3})Y^2 - Y^2 Z+ e^{t_2+t_3}Y + (e^{t_1+t_2+t_3} + e^{t_3})YZ
 - e^{t_1+t_2+2t_3}Z=0.
\ea

\subsection{Eigenvalue distribution}

The exact eigenvalue distribution
can be obtained by solving (\ref{specurve}) for $v(z)$
via $Y=\exp(z+v)$
and by computing the eigenvalue density $\rho\propto v_+(z)-v_-(z)$
along the cuts.
Here we apply force balance to derive the approximate distribution when
 \ba\label{param-region}
g_s n_I\gg 1,~~ g_s k_I\gg 1~~\hbox{ for all } I.
 \ea
Force balance is easier to understand intuitively.

\begin{figure}[htbb]
\begin{center}
\includegraphics[scale=.3]{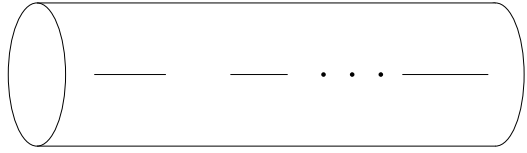}  
\caption{The eigenvalues are distributed along $m+1$ cuts
on the cylinder parametrized by $z$.
}
\label{S3-eigenvalues}
\end{center}
\end{figure}

We make the assumption, to be justified
a posteriori, that
\ba\label{assum}
u^{(I)}_i-u^{(J)}_j \gg 1~\hbox{for all }I,J,i,j\hbox{ such that }I<J.
\ea
Because the last term in (\ref{S3eom}) 
becomes constant we have
\be 
u_i^{(I)} =
g_s \sum_{j\neq i}\f2{1-e^{u^{(I)}_j-u^{(I)}_i}}
+g_s\left(\sum_{J=1}^m k_J-\sum_{J=1}^{I}n_J
+\sum_{J=I+1}^{m+1} n_J\right).\label{S3eom2}
\ee
We expect that when $g_s n_I$ is large,
the eigenvalues of $u^{(I)}$ spread over a large region,
allowing us to approximate the function
$1/(1-e^{x})$ in (\ref{S3eom2})
by a step function.
If we order the eigenvalues so that
$u^{(I)}_i<u^{(I)}_j$ for any $i<j$, 
it follows that
\ba
u^{(I)}_i=2g_s i
+g_s\left(\sum_{J=1}^m k_J-\sum_{J=1}^{I}n_J
+\sum_{J=I+1}^{m+1} n_J\right),~~i=1,\ldots,n_I.\label{u-dist}
\ea
Along the $I$-th cut that
has width $2g_s n_I$,
the eigenvalues of $u^{(I)}$ are
distributed uniformly.
The $I$-th and $I+1$ cuts are distance
$g_s k_I$ apart from each other.\footnote{
Since $u$
 is the holonomy along $\alpha$,
its eigenvalue distribution is
 different from the distribution (\ref{bkgd-holo}) of
 holonomy $\oint_{\beta}\Acal$. 
In particular the eigenvalues are quantized in unit of $2g_s$.
 It should be possible to physically explain (\ref{u-dist})
 using the fact that the matrix model
 captures the Wilson loop in a non-canonical framing \cite{Aganagic:2002wv}.
 } 
We can thus justify the approximations above
when $g_s n_I$ and $g_s k_I$ are all large.
See Figure \ref{S3-eigenvalues}.
As discussed above, this sheet is connected to other $m+1$ sheets
through the $m+1$ cuts as shown in Figure \ref{snaky-S3}(a).

 
\section{Bubbling Calabi-Yau for lens space
from a matrix model}\label{lenssection}
\subsection{Matrix model for a Wilson loop in lens space}
A simple generalization of
the topological A-model on $T^*S^3$ is 
the orbifold $X_p\equiv T^*(S^3/{\mathbb Z}_p)$ \cite{Aganagic:2002wv}. The particular orbifold action is such that $S^3/{\mathbb Z}_p$ is the 
lens space $L(p,1)$.
This  space is defined by the equation
\ba
|z_1|^2+|z_2|^2=1
\ea
for complex variables $z_1$ and $z_2$, together with identification
\ba
(z_1,z_2)\sim (e^{2\pi i/p} z_1, e^{-2\pi i/p}z_2).
\ea
We  study the  Wilson loop 
\ba
W_R=\Tr_R P e^{\oint A}
\ea
along a circle that is the generator of the fundamental group.
We assume that the circle is the unknot.

The $U(N)$ Chern-Simons theory on $L(p,1)$
has many vacua.
Since the  equation of motion is
solved by a flat connection, the vacua are
in one-to-one correspondence
with the $N$-dimensional representations of $\pi_1(S^3/\Z_p)=\Z_p$.
The group $\Z_p$ is abelian, so any such representation
is a sum of one-dimensional ones.
A one-dimensional representation is specified by an integer
$a=1,\ldots,p$.
Thus a vacuum is specified by a partition of $N$:
\ba
N=N_1+N_2+\ldots+N_p.
\ea
Here $N_a$ is the number of times the $a$-th irrep appears.
The contribution of this vacuum to the partition function
is given by
\be
{\cal Z}_p= \int  \prod_{i=1}^{N} du_{i}  \prod_{i<j}\left(2\sinh 
\frac{u_i -u_{j}}{2}\right)^2 
\exp\left(-\frac{p}{2g_s}\sum_{i} u_i^2+\frac{2\pi i}{g_s}\sum n_i u_i
\right).
\label{LensZ}
 \ee
This matrix model was formulated in \cite{Marino:2002fk},
and was studied for example in \cite{Aganagic:2002wv,
Halmagyi:2003ze, Halmagyi:2003mm, Dolivet:2006ii}.

According to the prescription in \cite{Aganagic:2002wv}
(see also \cite{Bouchard:2007ys}),
the contribution from this vacuum to the Wilson loop vev
is given by
\be
\langle W_R\rangle_p
\hspace{-1mm}
=
\hspace{-1mm}
 \int  \prod_{i=1}^{N} du_{i}  \prod_{i<j}\left(
2\sinh \frac{u_i -u_{j}}{2}\right)^2 
\exp\left(-\frac{p}{2g_s}\sum_{i} u_i^2+\frac{2\pi i}{g_s}\sum n_i u_i
\right)
\Tr_R \hspace{.5mm} {\rm diag}(e^{u_i}), \label{LensWp}
\ee
where $\vec{n}$ is a vector of integers 
\ba
\vec n=(\stackrel{N_1}{\overbrace{1,\ldots,1}},\stackrel{N_2}{\overbrace{2,\ldots,2}}
,\ldots, \stackrel{N_p}{\overbrace{p,\ldots,p}}).
\ea
The spectral curve for this matrix model can be derived \cite{Halmagyi:2003ze} and it agrees with the string theory prediction \cite{Halmagyi:2003mm}. 

For a large representation $R$ with large values of
$n_I$ and $k_I$, we expect a large backreaction
of fields to the Wilson loop insertion.
We propose that the gauge field path-integral
has now more saddle points.
Each saddle point 
specified by $(N_a)$ before insertion
splits into many each of which is specified
by non-negative integers $(N_{Ia})$ satisfying the constraints
\be
\sum_{a=1}^p N_{Ia}=n_I,\ \ \ \sum_{I=1}^{m+1} N_{Ia}=N_a.
\label{Wrpconstraints}
\ee

We will argue that the contribution to the Wilson loop vev  
from the saddle point specified by 
the $(N_{I a })$ is given by the multi-matrix model
\ba 
&&\langle W_R\rangle_{p}^{(N_{I a})}
\nn\\
&=&
\hspace{-3mm}
\int  
\prod_{I=1}^{m+1}
\prod_{a=1}^p d_H u^{(Ia)} 
 \exp\left(-\f{p}{2g_s}\Tr (u^{(Ia)})^2  
+ \left(L_I+\f{2\pi i}{g_s}a\right) \Tr u^{(Ia)}\right)
\label{lens-model}
\\
&&\times
\prod_{I,a<b}
\det\left(
2\sinh\f{u^{(Ia)}\otimes 1-1\otimes u^{(Ib)}}2 \right)^2
\hspace{-3mm}
\prod_{I<J,a,b} 
\hspace{-1mm}
\det\left(
2\sinh\f{u^{(Ia)}\otimes 1-1\otimes u^{(Jb)}}2 \right)^2. 
\nn
\ea
Wilson loops in the lens space matrix model have also been considered in the interesting recent work \cite{Bouchard:2007ys} and 
it would be of interest to apply their methods
to the spectral curve in this paper. 


\subsection{Physical derivation of the matrix model}
We now derive the matrix model from a D-brane configuration that realizes the
Wilson loop in a lens space.  

Let us recall that $X_p$ is a $\Z_p$ orbifold of the
deformed conifold given by (\ref{def-con-eq}).
The orbifold action is generated by
\ba
(z_1,z_2, z_3, z_4)\ra (e^{-2\pi i/p} z_1, e^{2\pi i/p} z_2, 
e^{2\pi i/p} z_3, e^{-2\pi i/p} z_4), 
\ea
and the $\Z_p$ action on the $S^3$ given by $z_2=z_1^\ast,~z_4=-z_3^\ast$
(so $|z_3|^2+|z_1|^2=\mu$)
defines the lens space $L(p,1)=S^3/\Z_p$.
Since the $\Z_p$ only acts on the phases,
$X_p$ is still a fibration of $T^2\times \R$ over $\R^3$.
Let us redefine $\alpha$ to be the 1-cycle 
corresponding to the generator of the fundamental group, 
and  $\beta$ the 1-cycle given by the $2\pi$
phase rotation of $z_3$.
We use the axes of the two cylinders (given by
$z_1 z_2={\rm const.},~z_3 z_4={\rm const.}$) and the ${\rm Re}(w)$
direction as the base $\R^3$.
The cycle $\beta$ degenerates at $w=\mu$ and so does $\beta'\equiv -p\alpha+\beta$
 at $w=0$.
The cycle $\alpha$ never degenerates.

\begin{figure}[htbb]
\begin{center}
\begin{tabular}{ccc}
\psfrag{cyc1}{$\beta$}
\psfrag{cyc2}{$-p\alpha+\beta$}
\includegraphics[scale=.45]{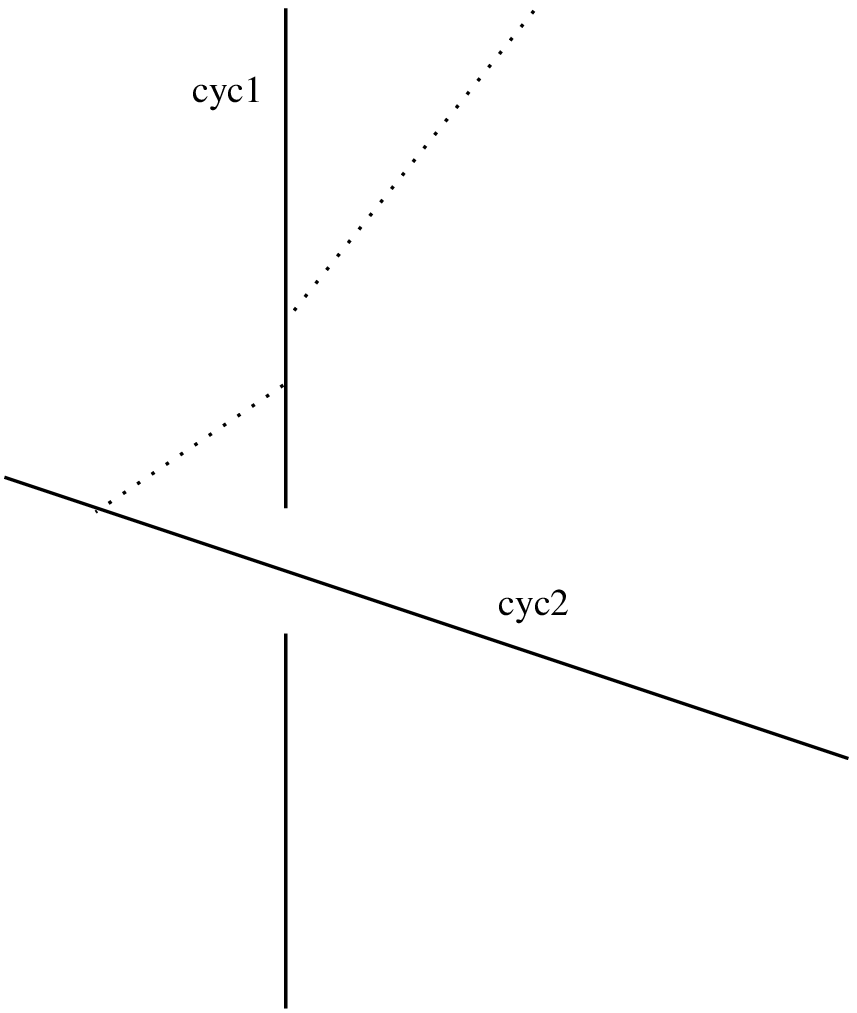} 
&
\psfrag{n1}{$n_1$}
\psfrag{n2}{$n_2$}
\psfrag{nm}{$n_m$}
\psfrag{nm+1}{$n_{m+1}$}
\includegraphics[scale=.45]{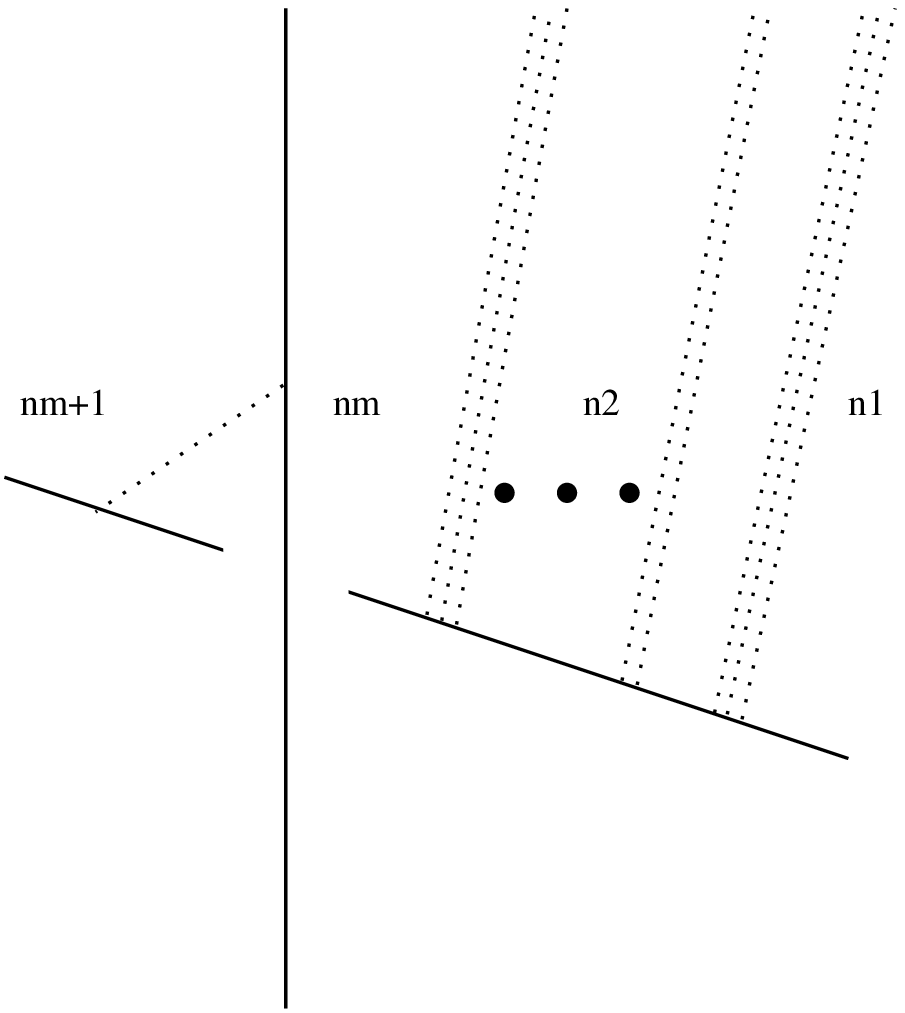} 
&
\psfrag{n1}{$n_1$}
\psfrag{n2}{$n_2$}
\psfrag{nm}{$n_m$}
\psfrag{nm+1}{$n_{m+1}$}
\includegraphics[scale=.45]{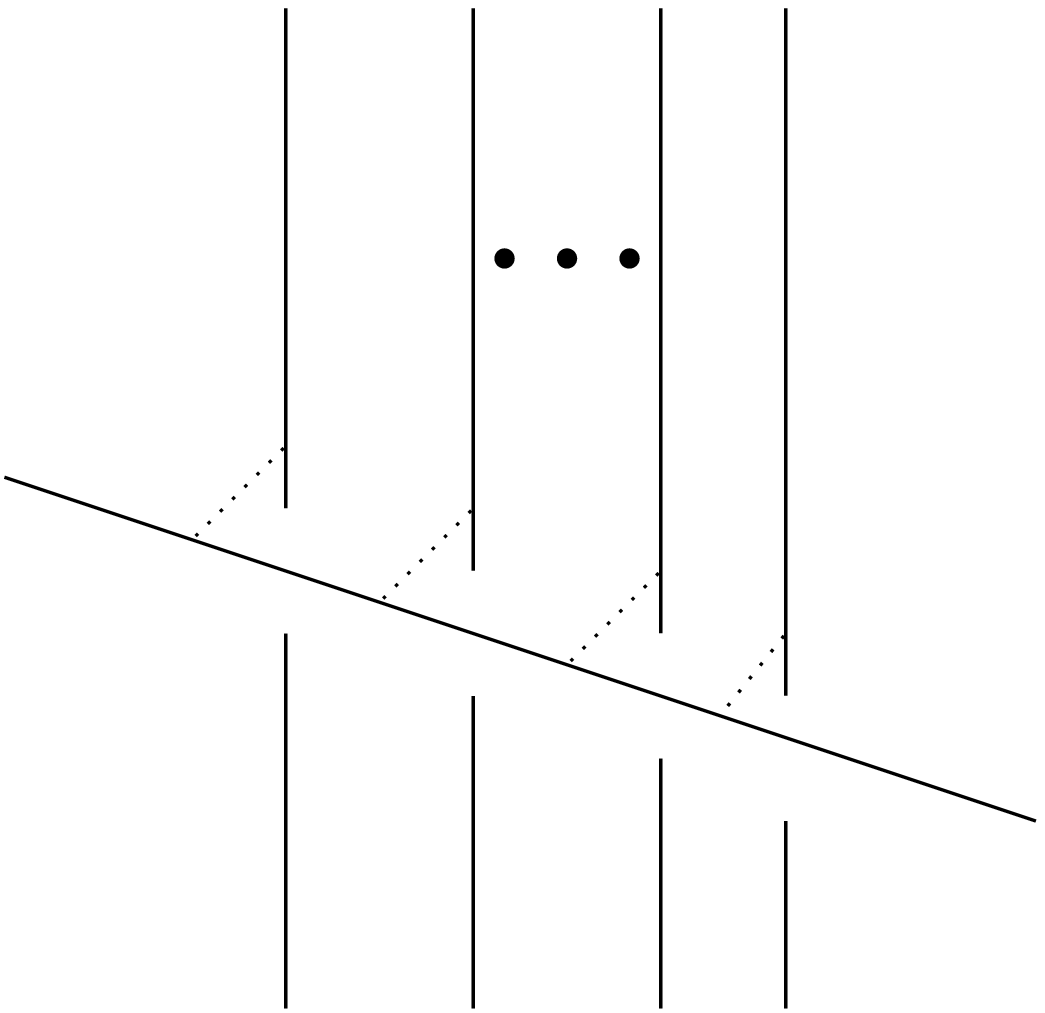}  
\\
(a)&(b)&(c)
\end{tabular}
\caption{(a) The cycle $\beta$ degenerates
along the vertical line while $-p\alpha+\beta$ degenerates along 
the other line.
If a linear combination $q\alpha+r\beta$ degenerates,
it does so
along a line in the $(q,r)$ direction.
(b) $P$ non-compact branes are distributed
along the line where $-p\alpha+\beta$ degenerates.
(c) There are $m+1$ copies of $S^3/\Z_p$.
}
\label{lens-cot}
\end{center}
\end{figure}

We engineer $U(N)$ Chern-Simons theory by wrapping
$N$ D-branes on the $S^3/\Z_p$.
To insert a Wilson loop along the knot $\alpha$,
we consider $P$ D-branes that wrap the non-compact cycle
$L=\R^2\times S^1$ in which $\beta$ is contractible.
See Figure \ref{lens-cot}(a).
The boundary condition
$\langle R|$ 
on the $P$ branes
picks out the Wilson loop insertion in representation $R$,
as explained in \cite{Gomis:2006mv}.
The boundary condition 
induces holonomy
\ba
\oint_{\beta=p\alpha+\beta'}\Acal=
{\rm diag}\left(g_s\left(R_i-i+\half(P+N+1)\right)\right)_{i=1}^P
\ea
along the contractible cycle $\beta=p\alpha+\beta'$.
By fibering the $T^2$ over a semi-infinite line ending on
the locus where $\beta'$ degenerates,
we obtain a 3-manifold in which $\beta$ is non-contractible.
We can consider a configuration of D-branes 
wrapping this 3-manifold.
Is the configuration equivalent to the one we started with,
as in the $S^3$ case?
We assume it is, and we will see evidence below.
The basic nontrivial cycle in
the new 3-manifold is $\alpha$, and the holonomy along it is given by
\ba
\int_{\alpha}\Acal=\oo p {\rm diag}\left(g_s\left(R_i-i+\half(P+N+1)
\right)\right)_{i=1}^P
\ea
because $\beta'$ is contractible.
See Figure \ref{lens-cot}(b).

As in the $S^3$ case, it is natural to split  the $P$ non-compact branes
into $m$ stacks with the $I$-th stack containing $n_I$ branes.
We can now replace $X_p$ by the $\Z_p$ orbifold of
the large $N$ dual geometry given by the equations (\ref{new-geom-eq}).
This is possible because (\ref{new-geom-eq})
are invariant under the orbifold action.
The non-compact branes are now replaced by compact ones
wrapping copies of lens space $S^3/\Z_p$.
Thus we reach the desired system of D-branes,
whose world-volume theory is
$m+1$ copies of Chern-Simons theory on lens space $S^3/\Z_p$,
interacting via Ooguri-Vafa operators.
The system is shown in Figure \ref{lens-cot}(c).

To write down the matrix model, we need to choose the vacuum of the theory.
We have a $U(n_I)$ Chern-Simons
theory on the $I$-th lens space.
As reviewed in the previous subsection,
the theory has many vacua corresponding
to the choice of a flat connection.
Let us choose the
vacuum specified by the partition $n_I=\sum_a N_{Ia}$.
Then according to the prescriptions in \cite{Aganagic:2002wv}, 
the contribution to the Wilson loop vev from this vacuum is given by
\ba
\langle W_R\rangle_{S^3/\Z_p}^{(N_{I a})}
&\sim&
\hspace{-1mm}
\int  \prod_{I,a,i} d_H u^{(Ia)} 
 \prod_{I,a<b}
\det\left(
2\sinh\f{u^{(Ia)}\otimes 1-1\otimes u^{(Ib)}}2 \right)^2
\hspace{-3mm}
\nn\\
&&\times
\prod_{I<J,a,b} 
\det\left(
2\sinh\f{(u^{(Ia)}+a_I/p)\otimes 1- 1\otimes (u^{(Jb)}+a_J/p)}2
\right)
\nn \\
&& \times
 \exp\left(-\f{p}{2g_s}\Tr (u^{(Ia)})^2  
+ \f{2\pi i}{g_s}a \Tr u^{(Ia)}\right). 
\ea
By redefining the variables as $u^{(Ia)}\ra u^{(Ia)}-g_s L_I/p
=u^{(Ia)}-a_I/p+(I$-independent),
we obtain (\ref{lens-model}).
It is remarkable that 
we get the holonomy $a_I/p$, including the factor of $1/p$,
which is necessary to be consistent with the algebraic derivation.
The success gives us confidence in the assumption we made above.

This brane construction makes it clear what the dual bubbling geometry
should be.
It should be the toric Calabi-Yau shown in Figure \ref{bub-geom}(b),
where all the copies of lens space have undergone
geometric transition.
This proposal will be confirmed in subsection 
\ref{lens-spec-sec} by deriving the
spectral curve of the matrix model, and showing
that it is the mirror of the toric Calabi-Yau.


\subsection{Algebraic derivation of the matrix model}

The vector of integers $\vec{n}$ in (\ref{LensZ}) breaks the $U(N)$ invariance down to the product subgroup $\times_a U(N_a)$ and subsequently the $\Scal_N$ symmetry to $\Scal^\prime =\times_{a} \Scal_{N_{a}}$. Nonetheless the Wilson loop is in the representation $R$ of $U(N)$ and as such we cannot immediately apply all the steps we used to solve the $S^3$ case in section \ref{alg-der-S3}. The workaround is to consider a generating function of matrix integrals, one term of which will correspond to the Wilson loop vev $\langle W_R \rangle_p^{(N_{I,a})}$. This generating function will have $\Scal_N$ symmetry and thus we need only the technology used in section \ref{alg-der-S3} to solve this case as well.

So we will consider
the generating function with variables $z_1,\ldots,z_p$
\ba
\hspace{-8mm}&&W_{R,p}(z_a)\nn\\
\hspace{-8mm}&=& \!\!\!\int \!\! \oo{N!}
\prod_{i=1}^{N} du_{i}  \prod_{i<j}\left(2\sinh 
\frac{u_i -u_{j}}{2}\right)^2\!
e^{ -\frac{p}{2g_s}\sum_{i} u_i^2} 
 \left(\prod_{i=1}^N\sum_{a=1}^{p}
e^{\f{2\pi i}{g_s}a u_i}z_a\right) \Tr_R
\hspace{.5mm} {\rm diag}(e^{u_i}). \label{LensWRgen}
\ea
The coefficient of $\prod_{a} z_{a}^{N_a}$ in (\ref{LensWRgen}) is $\langle W_R \rangle_p$. 
Since all the $u_i$ are dummy variables
on the same footing, 
we can straightforwardly repeat the analysis of section \ref{alg-der-S3} to arrive at
\ba
W_{R,p}(z_a)&=& \int\prod_{I=1}^{m+1}\left(\oo{n_I!}\prod_{i=1}^{n_I} du^{(I)}_i
 \prod_{i<j} \left(2\sinh \f{u^{(I)}_i-u^{(I)}_j}2\right)^2
e^{-\f{p}{2g_s} \sum_i (u^{(I)}_i)^2}
e^{L_I\sum_i u^{(I)}_{i}}
\right) \nn\\
&& \times \prod_I\prod_i 
\left(\sum_a e^{\f{2\pi i }{g_s} a u^{(I)}_i}z_a\right)
\prod_{I<J}\prod_{i,j}
\left(2\sinh \f{u^{(I)}_i-u^{(J)}_j}2\right),
\ea
where the eigenvalues $(u_i)$ have been divided into $m$ groups $(u_i^{(I)})$ ($m$ is again the number of groups of rows in $R$).  To understand the coefficient of $\prod z_a^{N_a}$ it is best to divide up the eigenvalues $(u_i)$ into $(u_i^{(a)})$ and also $(u_{i}^{(I,a)})$ such that
\ba
(u_i^{(I)})&=&\bigsqcup_{a=1}^p (u_{i}^{(Ia)}),\ \ \ (u_{i}^{(a)})=\bigsqcup_{I=1}^{m+1} (u_i^{(Ia)}), \nn \\
(u_i)&=&\bigsqcup_{I=1}^{m+1} (u_{i}^{(I)})=\bigsqcup_{a=1}^p (u_i^{(a)}). 
\ea
So clearly we have the constraints (\ref{Wrpconstraints})
and for each choice of non-negative integers $(N_{Ia})$ which satisfies these constraints, we have the following contribution to $\langle W_R \rangle_p$:
\ba
&&\langle W_R\rangle_p^{(N_{Ia})} \nn\\
&=& \hspace{-2mm}
\int \prod_I\Bigg[
\prod_{a,i}\frac{du^{(Ia)}_i}{N_{Ia}!}  
\prod_{a, i<j} \left(  2\sinh\f{u^{(Ia)}_i-u^{(Ia)}_j}2 \right)^2 \prod_{a<b,i,j}\left(
2\sinh\f{u^{(Ia)}_i-u^{(Ib)}_j}2 \right)^2
\label{Wrpterm}
\\
&&\times \exp\left(-\f{p}{2g_s}\sum_i (u^{(Ia)}_i)^2 +\sum_i (L_I+\f{2\pi i}{g_s}a) u^{(Ia)}_{i}\right)
\Bigg]
\prod_{I<J,a,b,i,j} 
\left(2\sinh \f{u^{(Ia)}_i-u^{(Jb)}_j}2\right).
\nn
\ea
This is the matrix model (\ref{lens-model}) in the eigenvalue basis.
We now solve the matrix model and derive its spectral curve.


\subsection{Spectral curve as bubbling geometry
for a  Wilson loop in lens space} \label{lens-spec-sec}

We now derive the spectral curve associated to (\ref{Wrpterm})
that captures the contribution of the particular vacuum
specified by the integers $(N_{Ia})$.
Since the gauge theory sums up such contributions,
the Wilson loop is actually dual to a sum over geometries.

The equation of motion
for $u^{(Ia)}_i$ that follows from (\ref{Wrpterm}) is
\ba
0&=&-pu_i^{(I,a)} +g_s L_I + 2\pi i a  
+g_s \sum_{j\neq i} \coth \frac{u_i^{(Ia)}- u_{j}^{(Ia)}}{2} 
+ g_s
\sum_{b\neq a, i}\coth \frac{u_{i}^{(Ia)}-u_j^{(Ib)}}{2} \nn \\
&&\ \ \ \ \ \ + \half g_s
\hspace{-2mm}
\sum_{J\neq I,b,i} 
\coth\f{u^{(Ia)}_{i}-u^{(Jb)}_{j}}2,
\label{lenseom}
\ea
and so we first define several resolvents
\be \label{resolvents-lens} 
v^{(Ia)}(z)=g_s \sum_{i=1}^{N_{Ia}}\f{e^{u^{(Ia)}_i}}{e^{u^{(Ia)}_i}-e^z},~~
v^{(I)}(z)=\sum_{a=1}^p v^{(Ia)}(z),~~
v(z)=\sum_{I=1}^{m+1}v^{(I)}(z).
\ee
In terms of these we can write (\ref{lenseom}) 
 as an equation on the $(Ia)$-cut:
\ba
p z+v_{\pm}(z)
=-v^{(I)}_\mp(z)+g_s\left(\sum_{J=I}^m k_J+\sum_{J=I}^{m+1}n_J\right)
+2\pi i a.
\label{lenseom2}
\ea

\begin{figure}[htbb]
\begin{center}
\begin{tabular}{ccc}
\psfrag{a}{$a$}
\psfrag{b}{$b$}
\includegraphics[scale=.6]{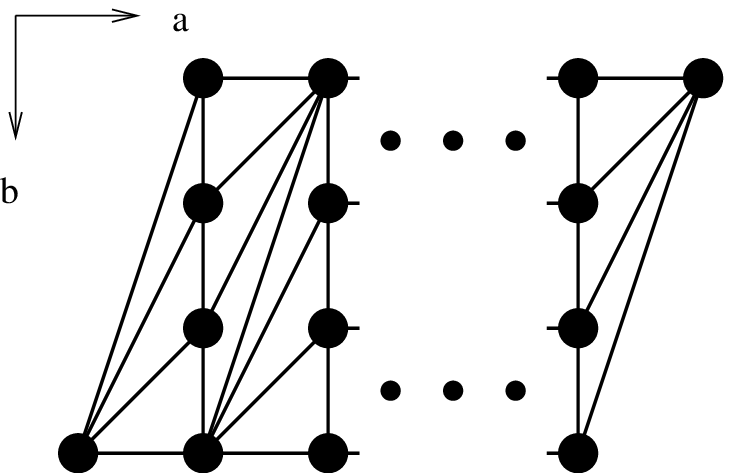} 
&
\hspace{1cm}
&
\includegraphics[scale=.3]{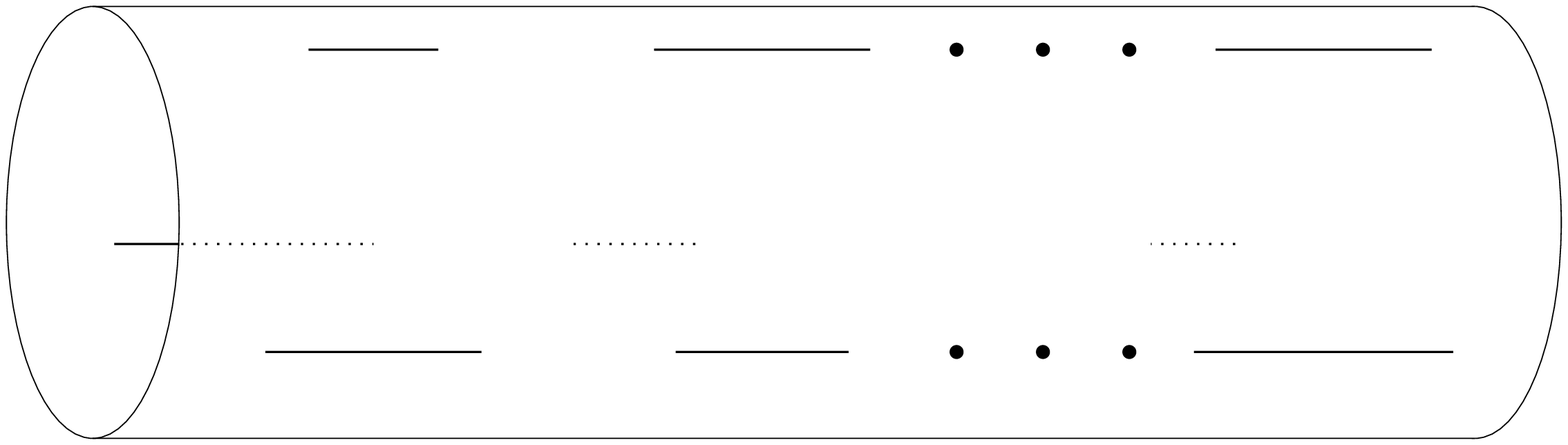}   
\\
(a)&&(b)
\end{tabular}
\caption{(a)
The vertices 
plot the monomials $Y^a Z^b$ in  (\ref{specurve-lens}).
By  connecting the vertices
by suitable edges, one obtains a graph that is dual to the toric
web in Figure \ref{bub-geom}(b).
(b) The eigenvalue distribution on the cylinder.
Here we chose $p=3$ for illustration.
}
\label{plot-lens}
\end{center}
\end{figure}

Following the same procedure as in section \ref{S3spect} we define some new variables\footnote{Despite identical nomenclature these variables are of course unrelated to those in section \ref{S3spect}.}
\ba
X_{0}&=&Z^p e^v , \nn \\
X_I&=&A_I e^{-v^{(I)}},\ \ \ \ I=1\ldots,m+1, 
\ea
where $Z=e^{z},\ A_I=\exp g_s\left(\sum_{J=I}^m k_J+\sum_{J=I}^{m+1}n_J\right)$.
Then the spectral curve is again given by (recalling once more that $(Y,Z)$ are ${\mathbb C}^*$ valued variables)
\be
f(Y,Z)=0 \label{specurve-lens}
\ee
where
\ba
f(Y,X_0,\ldots,X_{m+1})&=&\prod_{j=0}^{m+2}(Y-X_j)\nn \\
&=& \sum_{j=0}^{m+2} (-)^j Y^{m+2-j} E_{j}(X_0,\ldots,X_{m+1}).
\ea
The difference with section \ref{S3spect} lies in the asymptotics of 
the elementary symmetric polynomials,
from which we can determine their structure:
\ba
E_0(X_0,\ldots,X_{m+1})&=&1,~~
\nn\\
E_j(X_0,\ldots,X_{m+1})&=& \sum_{i=0}^{p}a_{j,i} Z^i
\hbox{ for } j=1,\ldots,m+1,
 \nn \\
E_{m+2}(X_0,\ldots,X_{m+1})&=&A_1\ldots A_{m+1} Z^p.
\ea
Some coefficients are easily determined:
\ba
a_{j,0}&=&\sum_{1\leq I_1<\cdots<I_j}
B_{I_1}\ldots B_{I_j}~\hbox{ for }j=1,\ldots,m+1,
\nn \\
a_{j,p}&=&
\sum_{1\leq J_1<\cdots <J_{j-1}} 
A_{J_1}\ldots A_{J_{j-1}}, ~\hbox{ for }j=2,\ldots,m+1,
~~
a_{1,p}=1,
\nn
\ea
where $B_I=\exp g_s\left(\sum_{J=I}^m k_J+\sum_{J=I+1}^{m+1}n_J\right)$.
The remaining $a_{j,i}$  
 are complex structure parameters
that are determined by demanding that $\oint v(z)dz=-2\pi i g_s N_{Ia}$
for the integral around the $(Ia)$-cut.
We can again write down the toric fan of the bubbling Calabi-Yau geometry directly from the spectral curve by plotting the monomials $Y^a Z^b$.
See Figure \ref{plot-lens}(a).

We see that this toric threefold is a daisy chain of lens spaces,
and the role of the complex structure deformations of the spectral curve is 
to desingularize each lens space. An interesting new feature of this geometry is the presence of nontrivial four-cycles.

\subsection{Eigenvalue distribution}
As in the $S^3$ case, when all $g_sn_I$ and $g_s k_I$ are large,
the interactions between $u^{(I)}$ and $u^{(J)}$ can be neglected.
The eigenvalue distribution for each $I$
is then that of a single lens space obtained in
\cite{Halmagyi:2003ze}.
According to \cite{Halmagyi:2003ze}, the eigenvalues of
$u^{(Ia)}$ are distributed
along a cut at ${\rm Im}(z)=2\pi a/p$ parallel to the real axis.
See Figure \ref{plot-lens}(b).
This sheet is connected to $m+1$ other sheets,
each through $p$ cuts.
The resulting topology is that obtained by fattening the
toric web in Figure \ref{bub-geom}(b).


\section*{Acknowledgments}
N.H. and T.O. acknowledge the hospitality of
the Simons Workshop at Stony Brook 
as well as the summer program at the Aspen Center for Physics,
where part of this work was done.
T.O. also thanks the Michigan Center for Theoretical Physics
and the Enrico Fermi Institute 
for hospitality.
The research of T.O. is supported in part by the NSF grants PHY-05-51164 and PHY-04-56556 and that of N.H. is supported by a Fermi-McCormick Fellowship and NSF Grants PHY-0094328 and PHY-0401814.

 \appendix

 \renewcommand{\theequation}{\Alph{section}\mbox{.}\arabic{equation}}

 \bigskip\bigskip
 \noindent {\LARGE \bf Appendix}


\section{Summary of Young tableau data}\label{Y-summary}
The Young tableau $R$ has $n_I$ rows of length $K_I$
such that $K_1>K_2>\ldots>K_m>K_{m+1}\equiv 0$.
It also has
$k_I$ columns of length $N_{m-I+1}$ such that
$N_1>N_2>\ldots>N_m$.
We also define $n_{m+1}\equiv N-\sum_{I=1}^m n_I,
 N_0\equiv N$, and $K_{m+1}\equiv 0$.
The integers $n_I,k_I,N_I$, and $K_I$ satisfy the relations
\ba
N_I= \sum_{J=1}^{m-I+1} n_J~ \hbox{ for }  I=0,1,\ldots,m ,
\ea
and
\ba
K_I= \sum_{J=I}^{m} k_J ~\hbox{ for } ~I=1,2,\ldots,m,~ K_{m+1}=0 .
\ea
See also Figure \ref{parametrization}.
We also denote by $P$ the number of rows in $R$,
so $P=N_1$.

Other useful sets of quantities are
\ba
L_I&=& \sum_{J=I}^{m} k_J -\half\sum_{J=1}^{I-1}n_J
+\half \sum_{J=I+1}^{m+1}n_J 
,\\
a_I&=&g_s\left(K_I-\left(n_1+\ldots+n_{I-1}+\half n_I\right)+\half(P+N)\right)
\nn\\
&=&g_s(L_I-L_{m+1}),
\ea
and
\ba
A_I&=&\exp g_s\left(\sum_{J=I}^m k_J+\sum_{J=I}^{m+1}n_J\right),  \label{AI}\\
B_I&=&\exp g_s\left(\sum_{J=I}^m k_J+\sum_{J=I+1}^{m+1}n_J\right).
\ea
\section{Area of the annulus diagrams}\label{appendix-area}

Here we explain the identification $a_I=g_s(L_I-L_{m+1})$
in subsection 
\ref{phys-der-S3}.

The $P$ non-compact D-branes with the boundary condition
$\langle R|$ has background holonomy \cite{Okuda:2007ai}
(gauge equivalent to the position relative to
the $N-P$ compact branes on $S^3$)
\ba
\oint_\beta\Acal={\rm diag}\left(g_s(R_i-i+\half(P+N+1))\right)_{i=1}^P
\label{bkgd-holo}
\ea
along the $\beta$ cycle.
When we split the $P$ non-compact branes
into $m$ stacks,
the average value of the holonomy in the $I$-th stack is
\ba
a_I=g_s\left(K_I-\left(n_1+\ldots+n_{I-1}+\half n_I\right)+\half(P+N)\right),
~~I=1,\ldots,m.
\label{aI}
\ea
Since this is the distance from the $S^3$,
it is natural to define $a_{m+1}\equiv 0$.
The parameters $a_I$ ($I=1,\ldots, m+1$) are then the positions of
$m+1$ copies of $S^3$ in the new geometry given by
(\ref{new-geom-eq}).
$a_I-a_{m+1}$ is the area of the annulus
between the $S^3$ and the $I$-th stack of non-compact branes.
See Figure \ref{physical1}(b).
Note that (\ref{aI}) can be written as
$a_I=g_s(L_I-L_{m+1})=g_s L_I
+$($I$-independent).

 \section{Alternative matrix models
 for a Wilson loop in \texorpdfstring{$S^3$}{S3}}\label{S3app} 

Here we discuss two alternative matrix models 
whose partition functions are the Wilson loop vev for
Chern-Simons on $S^3$.  The first is
\ba
\hspace{-1cm}&&\langle W_R\rangle
\nn\\
\hspace{-1cm}&=&
\int
 d_H u\hspace{1mm}  dU^{(1)} dU^{(2)}\ldots dU^{(m)}
e^{-\oo{2g_s}\Tr (u^2)}
(\det U^{(1)})^{k_{m}}(\det U^{(2)})^{k_{m-1}}
\ldots(\det U^{(m)})^{k_{1}}
\nn\\
\hspace{-1cm}&&\times \oo{\det(1-e^u\otimes U^{(1)}{}^\mo)}
\oo{\det(1-U^{(1)}\otimes U^{(2)}{}^\mo)}
\ldots
\oo{\det(1-U^{(m-1)}\otimes U^{(m)}{}^\mo)}.
\label{mat-mod-branes}
\ea
Here $U^{(I)}$ is an $N_I\times N_I$ unitary matrix.
The second is
\ba
\hspace{-1cm}&&
\left\langle
W_R\right\rangle
 \nn\\
\hspace{-1cm} &=&
\int d_H u  dU^{(1)} dU^{(2)}\ldots dU^{(m)}
e^{-\f{1}{2g_s}\Tr( u^2)}
(\det U^{(1)})^{n_1}(\det U^{(2)})^{n_2}\ldots(\det U^{(m)})^{n_m}
\nn\\
\hspace{-1cm}&&~
\times\det(1+e^u\otimes U^{(1)}{}^\mo) 
\oo{\det(1-U^{(1)}\otimes U^{(2)}{}^\mo)}
\ldots
\oo{\det(1-U^{(m-1)}\otimes U^{(m)}{}^\mo)},
\label{mat-mod-anti-branes}
\ea
for which $U^{(I)}$ is a $K_I\times K_I$ unitary matrix.

These models are obtained from (\ref{wilsonloop2})
by the same algebraic manipulations that led to similar multi-matrix models
for $\Ncal=4$ Yang-Mills in \cite{Okuda:2007kh}.

\subsection{Physical derivation} \label{alt-mat-phys}

Here we give a physical derivation 
of the matrix model
(\ref{mat-mod-branes}) from a D-brane configuration.

We begin with the configuration of $N$ compact 
and $P=N_1$ non-compact D-branes
(Figure \ref{physical1}(a)) that we discussed
in subsection (\ref{phys-der-S3}).
On the non-compact branes we impose the boundary condition $\langle R|$
to picks out the Wilson loop $W_R$ from
the annulus diagrams between the branes.

\begin{figure}[ht]
\begin{center}
\begin{tabular}{ccc}
\psfrag{F1R}{}
\psfrag{|Q2>}{$|Q^{(2)}\rangle$}
\psfrag{S1xS2}{$S^1\times S^2$}
\psfrag{N}{$N$}
\psfrag{N1}{$N_1$}
\psfrag{N2}{$N_2$}
\includegraphics[scale=.45]{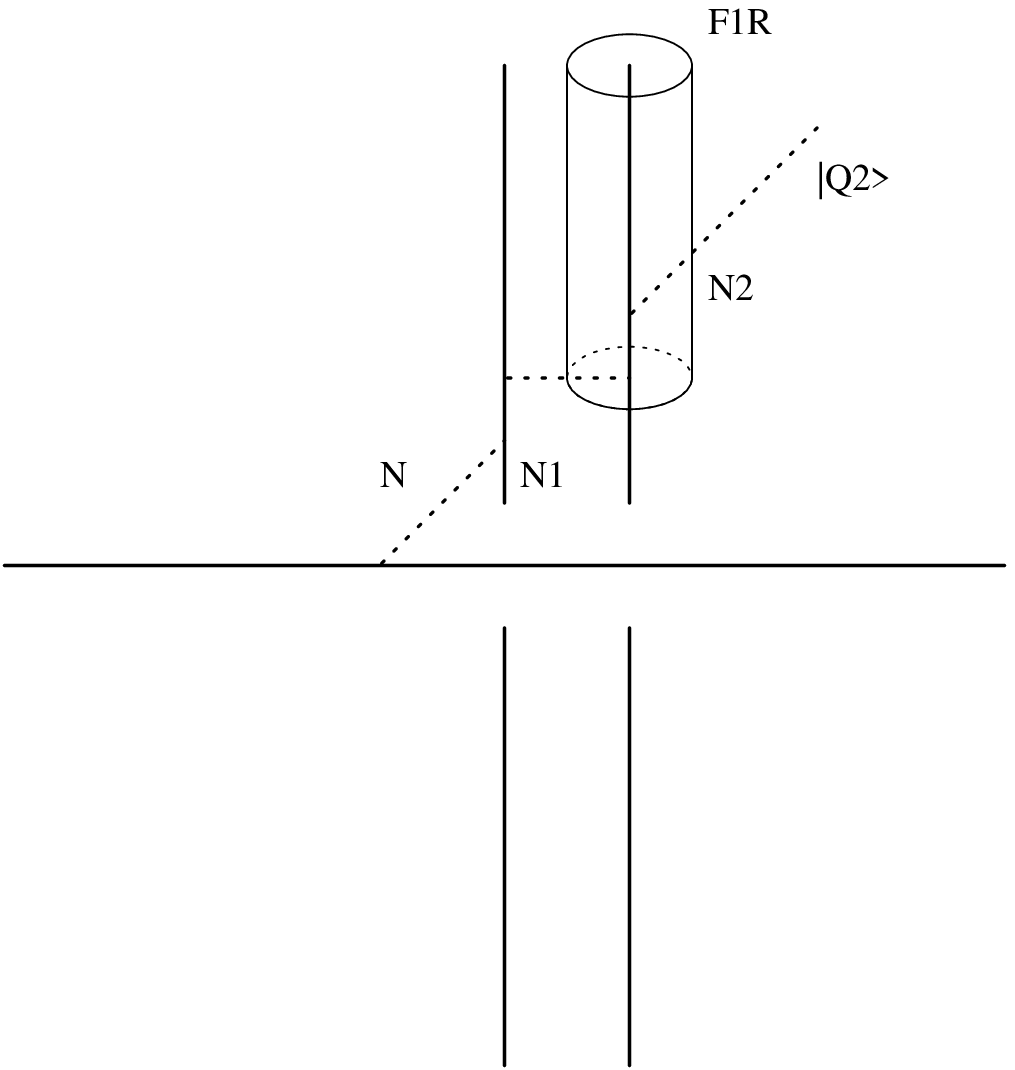}
&
\psfrag{N}{$N$}
\psfrag{N1}{$N_1$}
\psfrag{N2}{$N_2$}
\psfrag{Nm}{$N_m$} 
\includegraphics[scale=.45]{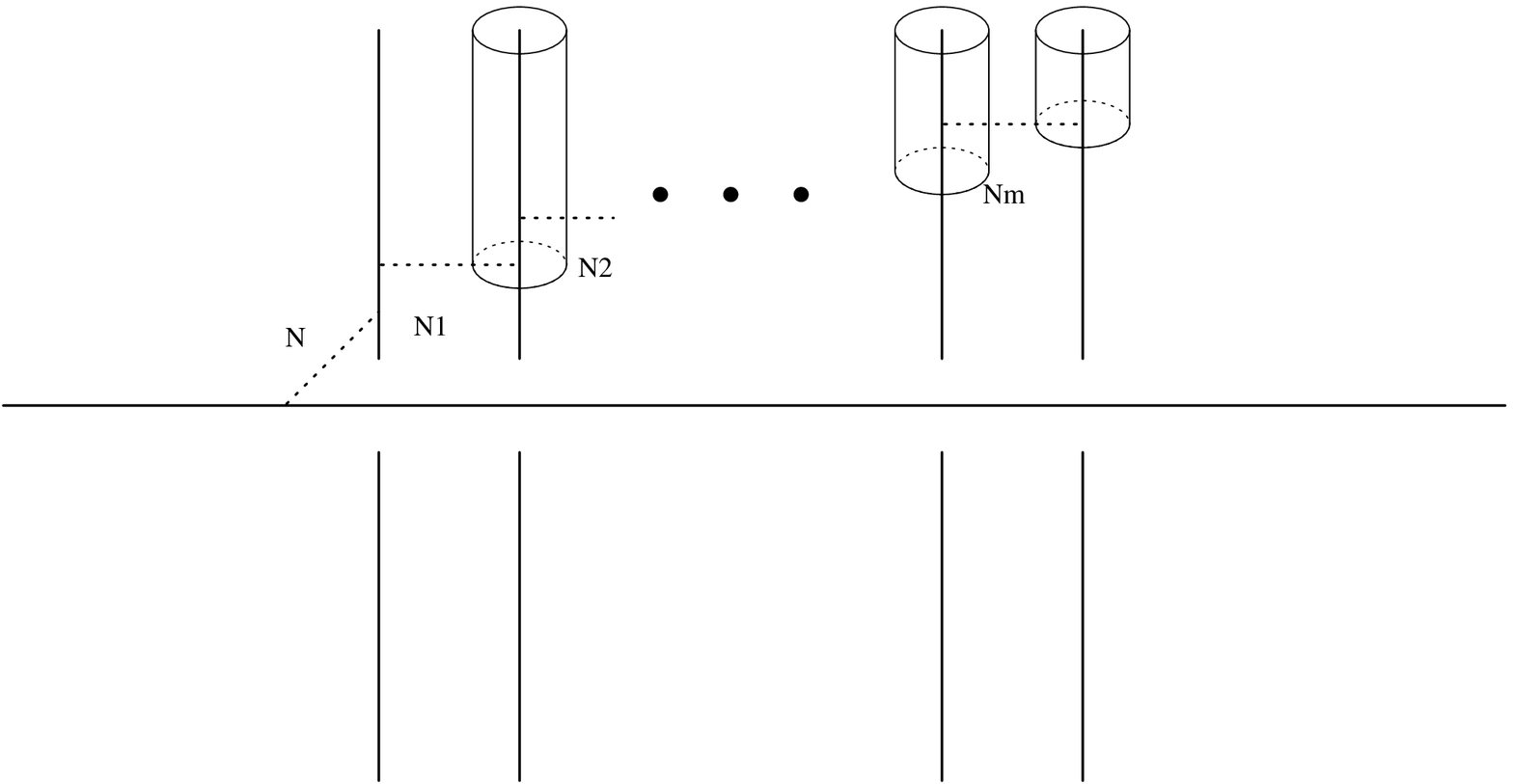}
\\
(a)&(b)
\end{tabular}
\end{center}
\caption{(a) The $P=N_1$ non-compact D-branes in Figure \ref{physical1} (a)
are compactified by modifying the Calabi-Yau geometry
without changing the topological string amplitudes.
The state $|R\rangle$ specifying the boundary condition 
is implemented by placing external string world-sheets
that insert the Wilson loop $\Tr_R \exp \oint A$.
(b) 
The geometry and the configuration of
D-branes and non-compact string world-sheets that
give rise to the multi-matrix model (\ref{mat-mod-branes}).
Each horizontal dashed line represents
D-branes wrapping a Lagrangian submanifold
of topology $S^1\times S^2$.
The cylinder ending on the $I$-th
dashed horizontal line
represents fundamental strings
in a configuration that inserts
a Wilson loop in the representation $A_{N_I}^{\otimes k_{m-I+1}}$
for $I=1,\ldots,m$.
}
\label{def-con-B}
\end{figure}

\begin{figure}[ht]
\begin{center}
\psfrag{R}{$R\equiv Q^{(1)}$}
\psfrag{Q2}{$Q^{(2)}$}
\psfrag{Qm-1}{$Q^{(m-1)}$}
\psfrag{Qm}{$Q^{(m)}$}
\psfrag{N1}{$N_1$}
\psfrag{N2}{$N_2$}
\psfrag{Nm-1}{$N_{m-1}$}
\psfrag{Nm}{$N_{m}$}
\includegraphics[width=120mm]{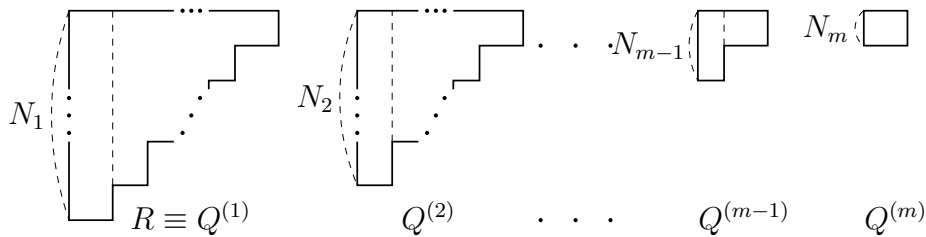}    
\caption{
A shrinking sequence of Young tableaux
$R\equiv Q^{(1)}\supset Q^{(2)}\supset\ldots\supset Q^{(m)}$.
}
\label{D-brane-split}
\end{center}
\end{figure}

We now consider a new configuration that realizes the Wilson loop insertion.
We  modify the geometry and introduce another locus on which
$\beta$ degenerates.
By fibering the $T^2$ over a line interval that connects
the two loci where $\beta$ degenerates, we get a cycle of topology
$S^1\times S^2$.
We wrap $N_1$ D-branes around this cycle while
placing external fundamental strings in an appropriate configuration.
This configuration of the fundamental strings is
that they insert the Wilson loop in the one-dimensional 
representation $A_{N_1}^{\otimes k_m}$ \cite{Okuda:2007ai}.\footnote{
$A_{N_1}$ is the rank $N_1$ totally anti-symmetric representation
of $U(N_1)$ and is one-dimensional.}
Additionally we place $N_2$ non-compact D-branes 
that end on the second locus where $\beta$ shrinks.
We choose the boundary condition to be $\langle Q^{(2)}|$,
where the Young tableau $Q^{(2)}$ is obtained from $R$ by removing
the first $k_m$ columns (Figure \ref{D-brane-split}).
The external strings and annulus diagrams from the non-compact branes
insert to the $S^1\times S^2$ branes the Wilson loop
\ba
\Tr_{A_{N_2}^{\otimes k_m}} e^{\oint A}\Tr_{Q^{(2)}}P e^{\oint A}
=\Tr_R P e^{\oint A}.
\ea
Since $S^1\times S^2$ is obtained by gluing two copies of solid torus
by identifying their boundaries, the path-integral
there reduces to the inner product.
Thus from the annulus diagrams between
the $S^3$ and $S^1\times S^2$, the path-integral picks out
the combination that inserts the Wilson loop $W_R$ into $S^3$.
See Figure \ref{def-con-B}(a).

We can repeat this process (Figure \ref{D-brane-split}) and show that
the following configuration is
equivalent to the Wilson loop insertion.
The total geometry 
is given by the same equation (\ref{new-geom-eq}) as in
subsection \ref{phys-der-S3}, with
one locus where $\alpha$ shrinks,
and $m+1$ parallel loci where $\beta$ shrinks.
$N$ D-branes wrap the original $S^3$.
We also wrap $N_I$ D-branes on the $S^1\times S^2$
between the $I$-th and $(I+1)$-th loci where $\beta$ shrinks.
Finally we place fundamental strings, along the $I$-th locus,
 that insert the Wilson loop
in the representation $A_{N_{I+1}}^{\otimes k_{m-I}}$
into the $I$-th $S^1\times S^2$.
See Figure \ref{def-con-B}(b).

Using the prescriptions in \cite{Aganagic:2002wv}, 
we obtain the matrix model (\ref{mat-mod-branes}) from this
D-brane configuration.
There is no Gaussian factor for the Chern-Simons on $S^1\times S^2$
since the path-integral is simply the inner product.
The external fundamental strings insert the determinant factors.

It is also easy to extend the derivation to
(\ref{mat-mod-anti-branes}), this time
using anti-branes instead of D-branes.
This explains the appearance of one determinant,
rather than the inverse of it, in (\ref{mat-mod-anti-branes}).

\subsection{Solving (\texorpdfstring{\ref{mat-mod-branes}}{mat-mod-branes})}
Now that we know the physical origin of 
the matrix model (\ref{mat-mod-branes}),
let us here solve it in the large $N$ limit.
In terms of the eigenvalues, the matrix model can be written as\footnote{
The quantities $u^{(I)}_i$ and $v^{(I)}$ 
in this subsection are not to be confused with
the quantities denoted by the same symbols in other parts of the paper.
}
\ba
\langle W_R\rangle
&\propto&
\int \prod_{i=1}^N du_i \prod_{I=1}^m \prod_{i=1}^{N_I} du^{(I)}_i
\exp\Bigg[
-\f{1}{2g_s}\sum_{i=1}^N u_i^2
+\sum_{I=1}^m k_{m-I+1}\sum_{i=1}^{N_I} u^{(I)}_i
\nn\\
&&+\sum_{i< j}\log \left(\sinh\f{u_i-u_j}2\right)^2
+\sum_{I=1}^m \sum_{i<j}\log  \left(\sinh\f{u^{(I)}_i-u^{(I)}_j}2\right)^2
\nn\\
&&-\sum_{i=1}^N \sum_{j=1}^{N_1}\log(1-e^{u_i-u^{(1)}_j})
-\sum_{I=1}^{m-1} \sum_{i=1}^{N_I}\sum_{j=1}^{N_{I+1}}
\log(1-e^{u^{(I)}_i-u^{(I+1)}_j})
\Bigg].
\ea
Proceeding as in subsection \ref{S3spect},
by defining the resolvents
\ba
v(z)&=&g_s\sum_{i=1}^N \f{e^{u_i}}{e^{u_i}-e^z},\nn\\
v^{(I)}(z)&=&g_s\sum_{i=1}^{N_I} \f{e^{u^{(I)}_i}}{e^{u^{(I)}_i}-e^z}
~\hbox{ for } I=1,\ldots,m,
\ea
we express the saddle point equations as
\ba
v_\pm(z)+z=-v_\mp(z)+v^{(1)}+g_s n_{m+1}
\ea
on the $u$-cuts and
\ba
-v^{(I-1)}(z)+v^{(I)}_\pm(z)=
-v^{(I)}_\mp(z)+v^{(I+1)}(z)
+g_s(k_{m-I+1}+n_{m-I+1})
\ea
on the $u^{(I)}$-cuts, for $I=1,\ldots,m.$
Note that we have defined $v^{(0)}\equiv v,~v^{(m+1)}\equiv 0$.
These equations state that the following quantities
are permuted as one goes through a cut:
\ba
X_0\equiv e^{v+z},
~~
X_{I}\equiv A_I e^{-v^{(m-I+1)}+v^{(m-I+2)}}
~~\hbox{ for }I=1,\ldots,m+1,
\label{newXI}
\ea
where $A_I$ is the familiar quantity defined in (\ref{AI}). 
The asymptotic behavior of $X_I$ as $z\ra \pm\infty$
is the same as that of $X_I$  in subsection \ref{S3spect}.
The rest of the analysis then goes exactly in the same way,
leading to the spectral curve (\ref{specurve}).
In particular $X_I$ here can be identified
with the quantity denoted by the same symbol there.
It was found there that $X_0$
has $m+1$ branch cuts,
while $X_I$ with $I=1,\ldots,m+1$ 
shares with $X_0$ just the $I$-th cut.
One can now show using (\ref{newXI}) that
$v^{(m-I+1)}(z)$ shares with $v(z)$
the first $I$ of these cuts,
and thus the $I$-th cut consists of the eigenvalues
of $u$, $u^{(1)}$,\ldots, and $u^{(m-I+1)}$.

How do we interpret the different kinds of eigenvalues
that lie along the same cut?
We believe that these eigenvalues form bound states
due to attractive forces, as explained in \cite{Okuda:2007kh} for
a matrix model that describes
a Wilson loop in $\Ncal=4$ super Yang-Mills.
The $I$-th cut has $n_I$ $(u$-$u^{(1)}$-\ldots-$u^{(m-I+1)}$)
bound states.

\subsection{Solving (\texorpdfstring{\ref{mat-mod-anti-branes}}{mat-mod-anti-branes})}
Let us also solve (\ref{mat-mod-anti-branes}),
which in terms of eigenvalues reads\footnote{
The quantities $u^{(I)}_i$ and $v^{(I)}$ 
in this subsection are not to be confused with
the quantities denoted by the same symbols in other parts of the paper.
}
\ba
\langle W_R\rangle
&\propto&
\int \prod_{i=1}^N du_i \prod_{I=1}^m \prod_{i=1}^{K_I} du^{(I)}_i
\exp\Bigg[
-\f{1}{2g_s}\sum_{i=1}^N u_i^2
+\sum_{I=1}^m n_I\sum_{i=1}^{K_I} u^{(I)}_i
\nn\\
&&+\sum_{i< j}\log \left(\sinh\f{u_i-u_j}2\right)^2
+\sum_{I=1}^m \sum_{i<j}\log  \left(\sinh\f{u^{(I)}_i-u^{(I)}_j}2\right)^2
\nn\\
&&+\sum_{i=1}^N \sum_{j=1}^{K_1}\log(1-e^{u_i-u^{(1)}_j})
-\sum_{I=1}^{m-1} \sum_{i=1}^{K_I}\sum_{j=1}^{K_{I+1}}
\log(1-e^{u^{(I)}_i-u^{(I+1)}_j})
\Bigg].
\ea
Again by defining the resolvents
\ba
v(z)&=&g_s\sum_{i=1}^N \f{e^{u_i}}{e^{u_i}-e^z},\nn\\
v^{(I)}(z)&=&g_s\sum_{i=1}^{K_I} \f{e^{u^{(I)}_i}}{e^{u^{(I)}_i}-e^z}
~\hbox{ for } I=1,\ldots,m,
\ea
the saddle point equations can be written as
\ba
\left(v(z)+z\right)_\pm =\left(-v(z)-v^{(1)}(z)+g_s (N+K_1)\right)_\mp
\ea
on the $u$-cuts,
\ba
\left(-v(z)-v^{(1)}(z)\right)_\pm=
\left(v^{(1)}(z)-v^{(2)}(z)
-g_s(n_1+k_1)\right)_\mp
\ea
on the $u^{(1)}$-cuts, and
\ba
\left(v^{(I-1)}(z)-v^{(I)}(z)\right)_\pm=
\left(v^{(I)}(z)-v^{(I+1)}(z)
-g_s(n_I+k_I)\right)_\mp
\ea
on the $u^{(I)}$-cuts for $I=2,\ldots,m$,
where we defined $v^{(m+1)}\equiv 0$.
From these equations we see that the following quantities
are permuted as one goes through a cut:
\ba
 X'_0\equiv e^{v+z},
~~
 X'_1\equiv A_1 e^{-v-v^{(1)}},
~~
 X'_{I}\equiv A_I e^{v^{(I-1)}-v^{(I)}}
~~\hbox{ for }I=2,\ldots,m+1,
\label{newXIt} 
\ea
where $A_I$ are defined in (\ref{AI}).
The asymptotic behavior of $X'_I$ as $z\ra +\infty$
is that of $X_I$,
but as $z\ra-\infty$, $X'_1$ behaves
like $X_{m+1}$, and $X'_{I}$ like $X_{I-1}$
for $I=2,\ldots,m+1$.
$X'_0$ and $X_0$ share the same asymptotics,
hence so do $E_j(X'_0,\ldots,X'_{m+1})$
and $E_j(X_0,\ldots,X_{m+1})$.
One concludes that the spectral curve of this model
is the one found in subsection  \ref{S3spect}.

What is the explanation of the difference
between $X'_I$ and $X_I$?
The functions $X_I$ are all holomorphic
 on the zero-th
sheet except on the $m+1$ cuts along the
real axis.
While 
\ba
(X'_0, X'_1,\ldots,X'_{m+1})=
(  X_0,   X_1,\ldots,X_{m+1})
\ea
for ${\rm Re}(z)$ that is positively large enough,
for negatively large  ${\rm Re}(z)$
we have
\ba
(X'_0, X'_1, X'_2,\ldots,X'_{m+1})=
(  X_0,   X_{m+1},X_1,\ldots,X_{m}). 
\ea
Thus $X'_I$ are not continuous,
and we believe that the discontinuities
arise due to the $v^{(I)}$-cuts ($I=1,\ldots,m$) that lie in 
the imaginary direction
as in \cite{Okuda:2007kh}.

\section{An improved matrix model for 
\texorpdfstring{$\Ncal=4$}{N=4} Yang-Mills}\label{YMsec}

This appendix is targeted at readers who are
interested in Wilson loops in the AdS/CFT context.

It is believed \cite{Drukker:2000rr,Erickson:2000af}
that the correlation functions of  circular
loops in $\Ncal=4$ Yang-Mills are
captured by the Gaussian matrix model.
The precise correspondence states 
in particular that
\ba
\left\langle \Tr_R P \exp \oint (A+\theta^i X^i ds)\right\rangle_{U(N)}
=
\oo Z \int dM \exp\left({-\f{2N}\lambda \Tr M^2}\right)
\Tr_R e^M. \label{mat-wil-U}
\ea
The left-hand side is the normalized expectation value
of the circular supersymmetric Wilson loop
in the Yang-Mills with gauge group $U(N)$.
The right-hand side is normalized by using
the partition function $Z$ which is
the integral without the insertion of $\Tr_R e^M$.
$dM$ is the standard hermitian matrix measure,
and $\lambda=g_{YM}^2 N$ is the 't Hooft coupling.
In the absence of operator insertions, the eigenvalues
are distributed according to the Wigner semi-circle law
in the large $N$ limit.

By applying the same algebraic manipulation as 
we did in subsection \ref{alg-der-S3},
we conclude that 
the vev of a circular Wilson loop is given by 
several Gaussian matrix integrals
correlated by interactions:
\ba
\langle W_R\rangle_{U(N)}
&=&
\oo Z
\int \prod_{I=1}^{g+1} dM^{(I)} e^{-\f{2N}\lambda \sum_I\Tr( M^{(I)})^2}
e^{K_I\Tr M^{(I)}}
\prod_{I<J}\det\f{(M^{(I)}\otimes 1-1\otimes M^{(J)})^2}{1-e^{-M^{(I)}}\otimes e^{M_J}}
\nn\\
&=&
\oo Z
\int \prod_{I=1}^{g+1}
\left(\oo{n_I!}
\prod_{i=1}^{n_I} dm^{(I)}_i 
 \prod_{1\leq i<j\leq n_I}(m^{(I)}_i-m^{(I)}_j)^2
e^{-\f{2N}\lambda \sum_i(m^{(I)}_i)^2} e^{K_I\sum_i m^{(I)}_i}
\right)
\nn\\
&&\times
\prod_{1\leq I<J\leq g+1}
\prod_{i=1}^{n_I}\prod_{j=1}^{n_J} 
\f{(m^{(I)}_i-m^{(J)}_j)^2}{1-e^{m^{(J)}_j-m^{(I)}_i}}.\label{YMmodel}
\ea
Here $M^{(I)}$ is an $n_I\times n_I$ hermitian matrix.
This is the direct analog of
the second expression in (\ref{WRfinal}).
We used the symbol $g$ to denote the number of blocks in $R$
as in \cite{D'Hoker:2007fq,Okuda:2007kh}, so $g=m$ in the notation 
of Figure \ref{parametrization}.

Using this multi-matrix model, it is remarkably easy
to obtain the eigenvalue distribution and
reproduce the Wilson loop vevs for the representations $R$
that are realized by a D3-brane \cite{Hartnoll:2006is, Yamaguchi:2007ps}, 
D5-brane \cite{Hartnoll:2006is, Yamaguchi:2006tq},
and bubbling geometry \cite{Okuda:2007kh}.
In particular, for an $R$ with large $g_{YM}^2 n_I$ and $g_{YM}^2 k_I$,
the gravitational dual is a smooth bubbling geometry.
Since $m^{(I)}_i$ is pulled to the right by the linear potential in
(\ref{YMmodel}) with coefficient $K_I$,
$m^{(I)}_i$ is much larger than $m^{(J)}_j$ if $I<J$.
Then the interaction between $M^{(I)}$ and $M^{(J)}$ can
be neglected. 
It then follows that for each $M^{(I)}$ the eigenvalues
are distributed around $\lambda K_I/4N$ according
to the semicircle law with half width $\sqrt{g_{YM}^2 n_I}$.

\bibliography{cs-mat}
\end{document}